\RequirePackage{fix-cm}
\documentclass[twocolumn]{svjour3}          
\smartqed  
\usepackage{graphicx}
\usepackage{booktabs}
\usepackage{amsmath}
\usepackage{cuted}
\usepackage{amssymb}
\usepackage{float}
\usepackage{subfigure}
\usepackage{tikz}
\usepackage{tikz-3dplot}
\usepackage{xcolor}
\usepackage{pgfplots}
\pgfplotsset{compat=1.18}
\usetikzlibrary{arrows, decorations.pathmorphing, positioning}
\usepackage{enumitem}
\usepackage[numbers,sort&compress]{natbib}
\usepackage{appendix}
\journalname{Eur. Phys. J. C}
\begin{document}

\title{Non-Equatorial Deflection of Light due to Kerr-Newman Black Hole: A Material Medium Approach
}

\titlerunning{Non-Equatorial Deflection of Light due to KNBH}        

\author{Saswati Roy \and 
        Shubham Kala \and
        Prasanjit Ghosh \and
        Hemwati Nandan \and
        Asoke K. Sen  
}


\institute{Saswati Roy \at
            Department of Physics, National Institute of Technology, Agartala, Tripura, 799046, India\\
            \email{sr.phy2011@yahoo.com}
            \and
            Shubham Kala \at
              The Institute of Mathematical Sciences\\
              C.I.T. Campus, Taramani, Chennai-600113\\
              Tamil Nadu, India\\
              \email{shubhamkala871@gmail.com} \and
            Prasanjit Ghosh \at
            Department of Physics, National Institute of Technology, Agartala, Tripura, 799046, India\\
            \email{prasanjitghosh699@gmail.com}
           \and
           Hemwati Nandan \at
              Department of Physics, Hemvati Nandan Bahuguna Garhwal Central University, Srinagar Garhwal, Uttarakhand 246174, India\\
              \email{hnandan@associates.iucaa.in}
            \and
            Asoke K. Sen \at
            Department of Physics, Assam University, Silchar, Assam 788011, India\\
            \email{asokesen@yahoo.com}
            }
           
\date{Received: date / Accepted: date}

\maketitle

\begin{abstract}
We explored the effect of space-time geometry on the trajectory of light rays in the context of a charged, rotating black hole. We derived an analytical expression for the deflection of light rays in Kerr-Newman space-time geometry, using a material medium approach, on non-equatorial plane. From this deflection angle expression it is evident that the charge and rotation of the black hole can affect the light rays' paths. For Kerr-Newman geometry, the deflection angle decreases with increasing charge when the rotation parameter is held constant. Conversely, for a constant charge, the deflection angle increases with the rotation parameter for prograde and decreases for retrograde trajectories. Applying both factors results in the deflection angle being lower than that of the Schwarzschild geometry. Non-equatorial study of the deflection angle reveals that it is maximum in the equatorial plane than in the pole. The frame-dragging effects in the Kerr-Newman field were taken into account to calculate the velocity of light rays, leading to the determination of the refractive index in this field geometry. This study concludes that, depending on the values of the rotation parameter and charge parameter, both prograde and retrograde trajectories coincide, result in that at some point the frame dragging effect is the same for prograde and retrograde motion. Also, the frame dragging effect increases towards poles than the equatorial plane, and this nontrivial nature results because of the interplay between charge, rotation parameter and rotation frequency of the black hole. 
\end{abstract}

%
%
\maketitle
%
%
\section{Introduction}
Einstein's General Relativity (GR) is one of the most revolutionary theories that fundamentally transformed our understanding of gravity and the structure of the universe.This theory posits that gravity arises from the curvature of spacetime caused by the presence of mass and energy like stars and planets warp the fabric of spacetime, and this curvature influences the motion of other objects, such as planets and light. One of the most famous predictions of GR is the bending of light around massive objects, known as gravitational lensing (GL), which has been observed and confirmed through astronomical observations.\\

The Kerr-Newman black hole (KNBH) extends the concept of black holes (BHs) beyond the simpler Schwar-zschild and Kerr solutions by incorporating both rotation and electric charge. The rotation of the KNBH introduces intriguing phenomena such as frame dragging, where the spacetime around the BH becomes twisted because of its rotation. This effect has implications for the behavior of matter and light near the BH. Furthermore, the electric charge of the KNBH alters its electromagnetic properties, influencing the behavior of charged particles and electromagnetic fields in its vicinity.\\

The study of light deflection in strong and weak gravitational fields began many decades ago. The confirmation of light deflection in 1919 \cite{dyson1920ix} was the starting point for developing gravitational lensing theory.  The deflection of the light ray can be calculated using different methods. The precise deflection of a light beam, when approaching a gravitational mass, is generally obtained by using the null geodesic through which the beam travels. An alternative method to calculate the deflection of a light beam near a gravitational mass is to assume that the light beam is traveling through a material medium, the refractive index of which is determined by the gravitational field. This method estimates the impact of gravitation on the light beam. This is known as the \textit{Material Medium Approach}.
In this approach, as the light ray propagates through a gravitational field, the effect of gravitation can be equivalently expressed as if the ray of light is passing through a material medium with some specific refractive index. For a centrally symmetric gravitational field, one can obtain a graded refractive index as a function of the radial coordinate  $r$ \cite{sen2010more}.  This method can also be extended to an axially symmetric field \cite{roy2015trajectory}, and in principle to other geometries \cite{roy2017deflection, roy2025deflection}. The concept of equivalent material medium was first pointed out by Tamm \cite{tamm1924electrodynamics} in 1924. Later, this concept was used by Balaz \cite{balazs1958effect} as early as 1958, and by Plebanski \cite{plebanski1960electromagnetic} in 1958. Atkinson \cite{atkinson1965light} investigated the allowed trajectories of light rays near a massive star and showed that the light ray is traveling both radially and tangentially to the field. And a general procedure for utilizing this concept was given by Felice \cite{de1971gravitational}.\\

From a scientific standpoint, the material medium approach offers a powerful and intuitive framework for modeling light propagation in curved spacetime, particularly in scenarios involving situations where a number of gravitational bodies are present, such as in binaries or multiple stellar systems, stellar clusters etc. It allows for the incorporation of medium-induced effects on photon trajectories and facilitates comparison with observational signatures, such as angular deflection and time delays. However, it is important to note that this approach does not claim observational superiority over the conventional null geodesic formalism. Both methods are theoretically consistent and independently capable of calculating light deflection. Moreover, the two approaches do not contradict one another; rather than being competing frameworks, they serve as complementary tools, with the choice between them guided by the specific requirements and context of the problem under investigation.
\begin{figure}[H]
    \centering
    \begin{subfigure}[]
        \centering 
        \begin{tikzpicture}[scale=0.7]

        \def\rBH{1.0}
        \def\rVac{2}
        \def\rMed{3}

        \fill[cyan!15] (0,0) circle (\rMed);         
        \fill[lightgray!30] (0,0) circle (\rVac);    
        \fill[white] (0,0) circle (\rBH);            
        \fill[black] (0,0) circle (\rBH);            

        \draw[very thick, red, ->] 
           (-3.5,1.2) 
           to[out=5, in=160] 
           (0.3,1.0) 
           to[out=-20, in=150] 
           (3.5,-0.5);

        \node at (0.6,0.3) [below left, text=white, font=\small\bfseries] {BH};
        \node[gray!50!black] at (0.1, -1.3) {\small Vacuum Region};
        \node[black!60!black] at (3.8, 0.8) {\small Material Medium};
        \node[red] at (-2.2,1) [left] {\small Incident Light Ray};
        \node[red] at (2.0,-0.7) [right] {\small Deflected Light Ray};

        \draw[dashed] (0,0) circle (\rVac);
        \draw[dashed] (0,0) circle (\rMed);

        \draw[->, thick, blue!60!black] (1.5, 2.5) -- (1.0, 1.7);
        \node[blue!60!black] at (2.3, 2.5) {\small Graded Refractive Index $n(r)$};

        \end{tikzpicture}
        \label{fig:schematic}
    \end{subfigure}
    \begin{subfigure}[]
        \centering
        \begin{tikzpicture}
        \begin{axis}[
            width=7cm,
            height=6.5cm,
            axis lines=left,
            xlabel={$r$},
            ylabel={$n(r)$},
            ymin=0.9, ymax=2.5,
            xmin=1, xmax=6,
            samples=100,
            domain=1.01:6,
            smooth,
            thick,
            every axis plot/.append style={blue!70!black},
            label style={font=\small},
            tick label style={font=\small},
            minor tick num=1,
        ]
        \addplot[blue, thick] {1 + 1/(x^2)};
        \end{axis}
        \end{tikzpicture}
    \end{subfigure}
     \caption{(a) Schematic view for the deflection of light due to a graded refractive index near a BH. Here, light bends because it follows the path of least optical distance in the refractive medium and vacuum acts as an effective medium influenced by the gravitational field. (b) The sample refractive index profile indicating the decline of refractive index, $n(r)$ with the distance \cite{roy2025deflection}.}
    \label{schematic}
\end{figure}
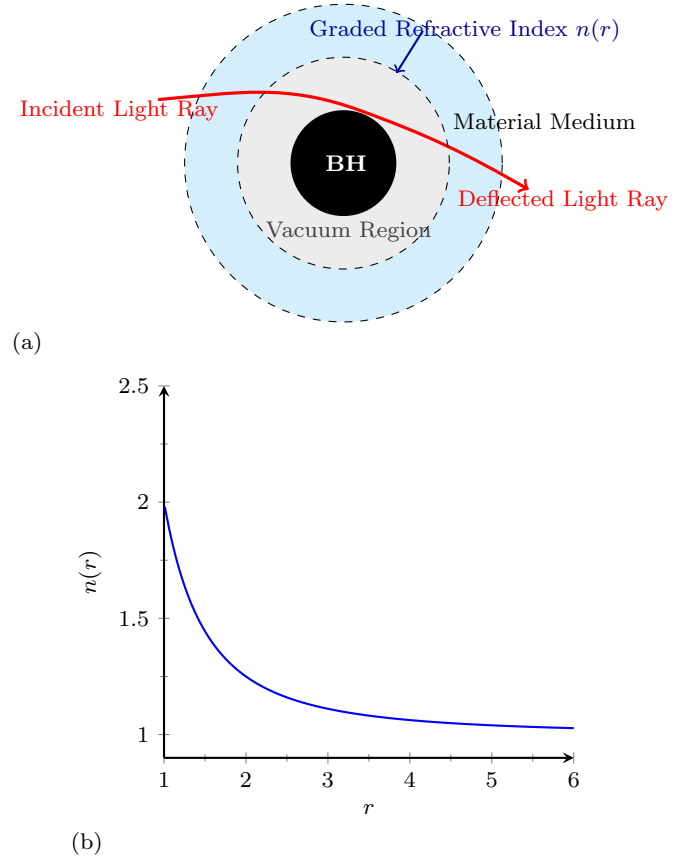
Several works have been done using null geodesic methods or the Hamilton Jacobi method for the Schwar-zschild black hole (SBH) \cite{misner1973gravitation, weinberg1972principles, schneider1992gravitational,virbhadra2000schwarzschild,keeton2005formalism,iyer2007light}, the Reissner Nordstr\"om black hole (RNBH) \cite{claudel2001geometry, eiroa2002reissner, bozza2002gravitational,bhadra2003gravitational, amore2006analytical,chakraborty2014solar}, the Kerr black hole (KBH) \cite{bray1986kerr,kopeikin2002gravitomagnetic,bozza2003quasiequatorial,bozza2005analytic,iyer2009light,aazami2011alensing,aazami2011blensing,chakraborty2017trajectory}, and the KNBH \cite{dadhich1977equatorial,de2000apparent,hasse2006morse,kraniotis2014gravitational,chakraborty2015light,hsiao2020equatorial,tinguely2020optical,hsieh2021gravitational,hsieh2021strong,kraniotis2022curvature,Kala:2022uog,chen2023radii}.  
Furthermore, the \textit{material medium approach} is also used to calculate the effect of a rotating body on the polarization of light \cite{tamm1924electrodynamics, balazs1958effect}, to study light scattering by gravitational field \cite{plebanski1960electromagnetic},  to obtain the deflection of an electromagnetic wave by gravitational field under different gravitating masses \cite{landau2013classical,de1971gravitational,mashhoon1973scattering, mashhoon1975influence,fischbach1980second}.
A group of researchers used the effective refractive index to calculate the gravitational time delay and trajectories of light rays in Schwarzschild geometry and showed that the opto- mechanical analogy of general relativity reproduces the equation of GR and matches with the classical equations \cite{evans1986f,evans1996optical}. The same work was continued also for the stationary gravitating mass to obtain the trajectory of the photon and obtained the Gravitational Faraday Rotation 
 \cite{ishihara1988gravitational,sereno2003gravitational, sereno2004weak,ye2008gravitational}. The exact frame-dragging frequency inside the rotating neutron star was obtained which depends on both the radial distance($r$) and the colatitude($\theta$) \cite{chakraborty2014dragging}. It is also observed that the normalized frame-dragging value is higher at the pole than at the equator for a particular pulsar. 

The conditions for equatorial circular geodesics in Kerr-Newman geometry were obtained in 1976 \cite{dadhich1977equatorial}. In 2000, it was observed that there do not exist any circular photon orbits of radius $r=r_\pm$, $r=r_g/2$, or $r=0$ in a Kerr-Newman space time and after studying apparent shadows of various KNBHs in the equatorial plane, it is recognized that the spherical symmetry of the shape is destroyed when rotation is present \cite{de2000apparent}. 
Hasse and Pelrick \cite{hasse2006morse} worked on the lensing of KNBH using Morse theory and showed that such a body formed an infinite number of images. Kraniotis \cite{kraniotis2014gravitational, kraniotis2022curvature} derived the analytic solutions of the lens equations to calculate the light deflection angle for KNBH and KNBH (anti) de Sitter. Chakraborty and Sen \cite{chakraborty2015light} obtained the light deflection angle for KNBH in the equatorial plane in terms of mass, spin and charge. In 2020,  the deflection angle due to the KNBH in terms of elliptic integrals was obtained \cite{hsiao2020equatorial}.
In 2021, Hsieh et al. \cite{hsieh2021strong,hsieh2021gravitational} studied strong gravitational lensing on the equatorial plane and gravitational time delay by KBH and KNBH.
In 2023, the equatorial and non-equatorial spherical photon orbits around KNBHs were also studied \cite{chen2023radii}.

Bozza \cite{bozza2003quasiequatorial} studied the quasi-equatorial lensing in the strong field limit for spinning BHs. Aazami et al. \cite{aazami2011alensing,aazami2011blensing} calculated the two components of the light bending angle, along the direction of the equatorial plane and perpendicular to the equatorial plane of a KBH in the quasi-equatorial regime. Chakraborty and Sen \cite{chakraborty2017trajectory} investigated the light ray to be travelling only slightly above the equatorial plane for KBH and obtained the off-equatorial light deflection angle for a slowly rotating body. Wang et al. \cite{Wang:2022ouq} derived geodesic trajectories in terms of elliptic integrals and Jacobian elliptic functions for both null and timelike geodesics, expressed as real functions of the Mino time around KNBH spacetimes. Chen et al. \cite{Chen:2024oyv} studied the extremal KNBH that permits a stable photon orbit exactly at the horizon due to the presence of the degenerate root in the radial potential.\\

Very recently, the graded refractive index approach was used to obtain the deflection angle and hence the photon trajectory under different gravitating masses such as SBH, KBH, RNBH, JNW, KSBH \cite{sen2010more, roy2015trajectory, roy2017deflection, roy2019study, roy2025deflection}. To obtain the deflection angle, we express the line element in an isotropic form and evaluate the refractive index to determine the trajectory of the light ray.

To the best of our knowledge, an off-equatorial or quasi-equatorial or non-equatorial plane has not been reported in the Kerr-Newman geometry to study the light bending angle or gravitational lensing. In this paper, we are the first to incorporate the non-equatorial deflection angle for KNBH using the material medium approach or the graded refractive index approach. 
In space we take the direction of the rotation of KNBH as the z-axis and the direction of the observer as the x-axis.  Also, consider that the light ray is approaching parallel to the x-axis from asymptotic infinity ($x$ or $r=-\infty$) toward a charged rotating gravitational mass and goes to infinity to the observer($x$ or $r = + \infty$) after a specific amount of deflection. From an observational point of view, we are not interested in other perspectives on the direction of light rays as these are insignificant. Thus, as the non-equatorial plane we are considering all the planes which are parallel to the equatorial plane and the x-axis.

The organization of the paper is as follows. Section 2 provides an overview and detailed discussion of the KNBH  spacetime, including the relevant parameters and horizon structure. Section 3 is devoted to the study of the photon sphere of the KNBH, considering two cases: the polar plane and the equatorial plane. In Section 4, we calculate the velocity of light in the KNBH field and determine the corresponding refractive index, followed by an in-depth analysis of the frame-dragging effect in a dedicated subsection. Section 5 focuses on the calculation of the deflection angle using the material medium approach at non-equatorial planes, accompanied by comprehensive graphical representations of all relevant parameters. In Section 6, we presented the comparison with the Kerr Sen black hole (KSBH) which arises from the low energy limit of heterotic string theory. Finally, we present our conclusions in Section 7.
\section{Kerr-Newman Black Hole}
The Kerr-Newman metric provides the precise solution to Einstein's Field Equation of GR for describing the gravitational field of a rotating, charged BH with stationary and axial symmetry. In the Boyer Lindquist form, in the ($ct$,$r$,$\theta$,$\phi$) coordinate system, the line element of the Kerr-Newman metric is \cite{newman1965metric,newman1965note,hsiao2020equatorial}
\begin{align}
 ds^2 =\ & \left(\frac{\Delta - \alpha^2 \sin^2\theta}{\Sigma^2} \right) c^2 dt^2 
- \frac{\Sigma^2}{\Delta} dr^2 
- \Sigma^2 d\theta^2 \nonumber \\
& - \left\{ (r^2 + \alpha^2)^2 - \alpha^2 \Delta \sin^2\theta \right\} 
\frac{\sin^2\theta}{\Sigma^2} d\phi^2 \nonumber \\
& + \frac{\alpha(r_g r - Q^2) \sin^2\theta}{\Sigma^2} c (dt\, d\phi + d\phi\, dt). 
\label{metric1}
\end{align} 
where
\begin{equation}
\begin{aligned}
\Sigma^2 &= r^2 + \alpha^2 \cos^2\theta, \\
\Delta   &= r^2 + \alpha^2 + Q^2 - r_g r.
\end{aligned}
\end{equation}
Here, $r_g=\frac{2GM}{c^2}$ is the Schwarzschild radius; $\alpha=\frac{J}{Mc}$ is the rotational parameter, where $J$ is the angular momentum of the gravitating body, $M$ is the total mass of the gravitating body and $c$ is the velocity of light; $Q^2=\frac{Ge^2}{4\pi\epsilon_0 c^4}$, where $e$ is the scalar charge, $\frac{1}{4\pi\epsilon_0}$ is the coulomb force constant, $G$ is the gravitational constant. It is evident that $Q$ and $r_g$ have the dimension of length.

After some straightforward algebraic manipulations, the above equation (\ref{metric1}) leads to a new simplified form as
\begin{align}
ds^2 =\ & \left(1 - \frac{r_g r - Q^2}{\Sigma^2} \right) c^2 dt^2 
- \frac{\Sigma^2}{\Delta} dr^2 
- \Sigma^2 d\theta^2 \nonumber \\
& - \left( r^2 + \alpha^2 + \frac{r_g r - Q^2}{\Sigma^2} \alpha^2 \sin^2\theta \right) \sin^2\theta\, d\phi^2 \nonumber \\
& + \frac{2\alpha(r_g r - Q^2) \sin^2\theta}{\Sigma^2} c\, dt\, d\phi.
\label{metric}
\end{align}
If we consider $Q=0$ the Kerr-Newman line element (\ref{metric}) reduces to the Reissner Nordstr\"om metric which represents the charged, non-rotating, or static gravitating mass. But if $\alpha=0$ the line element (\ref{metric}) is reduced to the Kerr line element, which represents the axially symmetric, uncharged gravitating mass. When both the rotational parameter $\alpha$ and the charge parameter $Q$ are absent, the line element (\ref{metric}) is reduced to the Schwarzschild metric, which is electrically neutral, non-rotating, or static gravitating mass. Like the Schwarzschild spacetime, the Kerr-Newman spacetime is also asymptotically flat. The coefficients of the line element are independent of $\phi$, which indicates that it is axially symmetric.

Singularity behavior for the metric components dependent on the coordinate systems. Kerr-Newman metric for any coordinate system reveals the presence of singularities, which are essentially the same as those of the Kerr metric. In Boyer-Lindquist coordinates, the Kerr-Newman metric has the curvature singularity at $\Sigma^2=0$, implying the curvature singularity at $r=0$ and $\theta=\pi/2$. However,  an additional singularity appears in terms of the metric component $g_{rr}$ which becomes singular when $\Delta=0$. 

Thus, the Kerr-Newman metric is regular at
\begin{equation}\label{horizon}
    r= r_\pm = \frac{1}{2}\left[r_g\pm \sqrt{r_g^2 - 4(\alpha^2 + Q^2)}\right], 
\end{equation}
or, 
\begin{equation}\label{horizon-}
    r= r_\pm = m \pm \sqrt{m^2 - (\alpha^2 + Q^2)},
\end{equation}
with the condition that $m^2>(\alpha^2 + Q^2)$ or $r_g^2>4(\alpha^2 + Q^2)$.
These surfaces are referred to as the outer $(r_+)$ and inner $(r_-)$ horizons; named the Event horizon and the Cauchy horizon, respectively. The region $r<r_+$ is referred to as the ‘interior’ of the BH.
The Kerr-Newman metric becomes extremal if $r_g^2 - 4(\alpha^2 + Q^2)=0$ and in such case, the inner and outer horizons coincide at $r=\frac{r_g}{2}$. If $r_g^2 - 4(\alpha^2 + Q^2)<0$, there is no real solution to $\Delta=0$  and no event horizon exists to hide the interior curvature singularity from exterior observers.\\
\begin{figure}[H] 
	\begin{center}       
        {\includegraphics[width=0.45\textwidth]{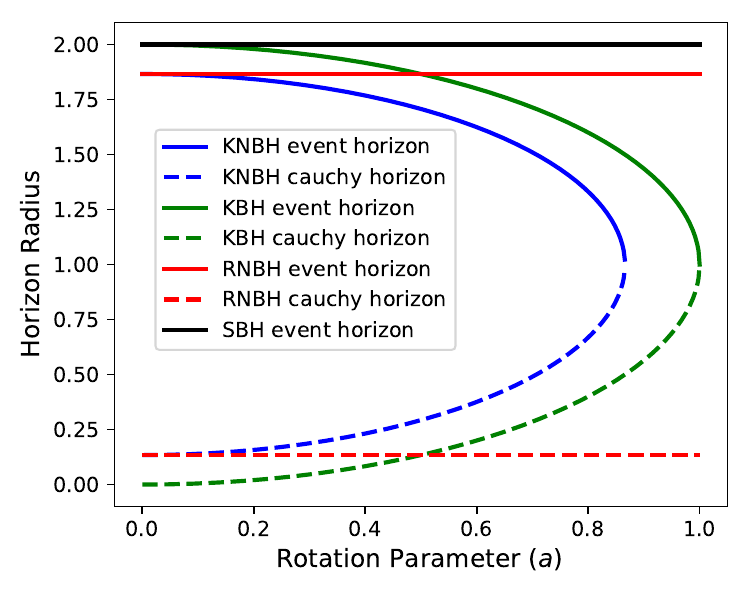}}
	\end{center}
	\caption{The variation of horizon radii with respect to the rotation parameter ($\alpha$) for KNBH as compared to other BHs solution in GR. Here we consider $M=1$, $u=0.4$ and $q=0.5$.} \label{FigHorizons}
\end{figure}

Fig.\ref{FigHorizons} illustrates the variation of the BH horizon radii as a function of the rotation parameter. The solid lines represent the event horizons, while the dashed lines denote the Cauchy horizons. For the SBH and RNBH, the horizon structure is independent of rotation, resulting in horizontal lines. In contrast, for the KNBH and KBH, the event horizon radius decreases with increasing charge, whereas the Cauchy horizon radius increases. Additionally, it is observed that the presence of charge generally reduces the size of the event horizon of BH. Among the cases considered, the SBH exhibits the largest event horizon radius, while the KNBH exhibits the smallest. This behavior highlights the combined influence of charge and rotation on the horizon structure in rotating charged spacetimes.

Another surface arises in Boyer-Lindquist coordinates by introducing a static world line, which has $r$, $\theta$ and $\phi$ as constant, so that the metric restricted to the world line. Such surfaces can be determined by considering $g_{tt}=0$ as
\begin{equation}
    r= r^E_\pm (\theta) = \frac{1}{2}\left[r_g\pm \sqrt{r_g^2 - 4(\alpha^2 \cos^2\theta + Q^2)}\right],
\end{equation}
called the stationary limit surface or the ergosurface.
The $r^E_+(\theta)$ is the outermost surface and the surface $r^E_-(\theta)$ lies inside the event horizon. 
The tangent to the static world line is time like for $r>r^E_+(\theta)$, while for $r<r^E_+(\theta)$, the world line is space like. The region between the ergosurface and the event horizon $r_+ < r < r^E_+(\theta) $  is referred as the ergosphere.
\section{The Photon Sphere}
A photon sphere is a region of space around a BH where gravity is so strong that photons (particles of light) can orbit the BH in stable circular paths. These paths are at a specific distance from the BH, known as the photon sphere radius. At this radius, the gravitational pull of the BH is strong enough to bend the path of light in such a way that it can orbit the BH indefinitely, assuming there are no other influences to disrupt its orbit.
The photon sphere is distinct from the event horizon, which is the boundary beyond which nothing, not even light, can escape the gravitational pull of the BH. While the event horizon marks the point of no return for objects falling into the BH, the photon sphere is a region where light can orbit the BH before ultimately being absorbed or escaping.\\
In order to obtain the photon sphere of a KNBH on a non-equatorial plane, one must solve a sextic equation, which is highly cumbersome to handle analytically. Therefore, we divide our analysis of the photon sphere into two special cases: the polar and equatorial plane, where analytical solutions can be obtained more feasibly. For this analysis, we closely follow the methodology proposed by Feng et al. \cite{Feng:2019zzn}.
\subsection{\textbf{Polar $(\theta=0)$}}
\begin{figure}
	\begin{center}
    \begin{subfigure}[]
     {\includegraphics[width=0.45\textwidth,height=0.321\textheight]{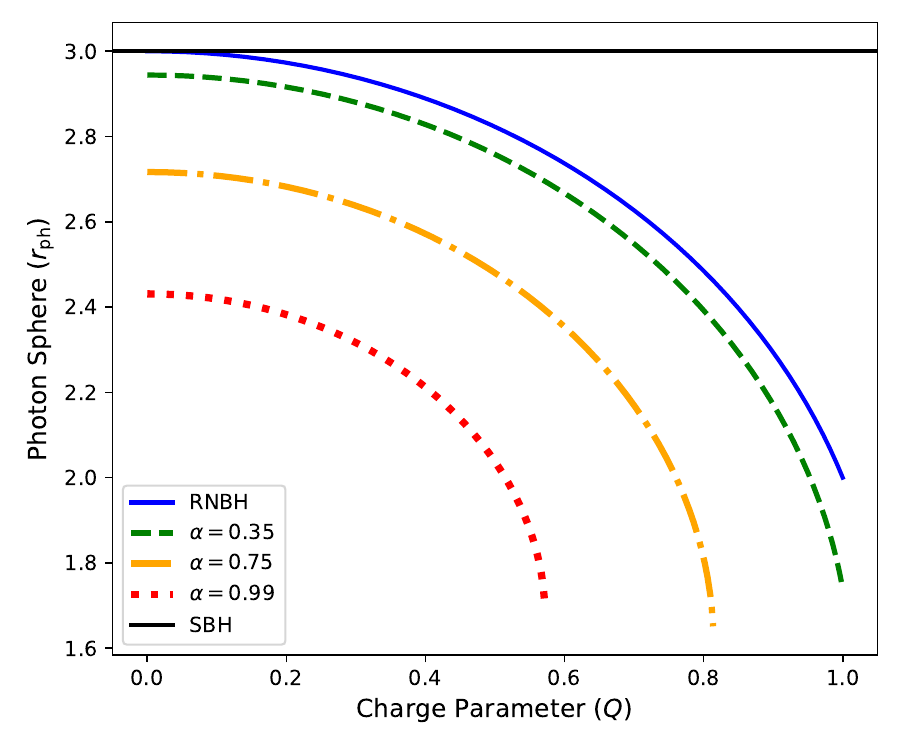}} 
    \end{subfigure}
     \begin{subfigure}[]
     {\includegraphics[width=0.45\textwidth,height=0.321\textheight]{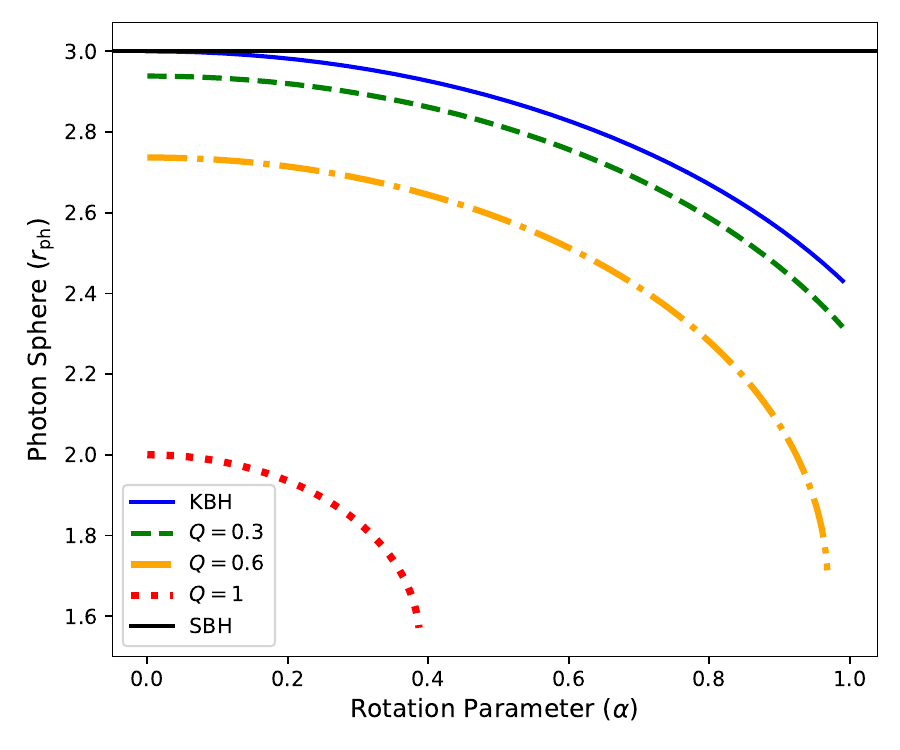}} 
    \end{subfigure}   
	\end{center}
	\caption{ The graphical representation of photon sphere radius at polar plane. Subfigure (a) corresponds to as a function of charge parameter with different values of rotation parameter, while subfigure (b) corresponds to as a function of rotation parameter with different values of charge parameter. Here we consider $M=1$ and $\theta=0$. } \label{PoleRPH}
\end{figure}

In this case, we have to deal with a cubic equation to obtain the photon sphere radius \cite{Feng:2019zzn},
\begin{equation}
 r_{\text{ph}}^3 - 3Mr_{\text{ph}}^2 + (\alpha^2 + 2Q^2) r_{\text{ph}} + M \alpha^2 = 0.
\end{equation}
The equation admits three possible roots: two positive and one negative. The negative root is unphysical, while among the two positive roots, one lies inside the BH horizon and the other lies outside. For the photon sphere to be physically relevant, it must lie outside the event horizon. Therefore, the physically meaningful photon sphere radius corresponds to the positive root located outside the horizon and is given by
\begin{equation}
\begin{split}
r_{\text{ph}} =\ & M + 2 \sqrt{M^2 - \frac{1}{3} \left(\alpha^2 + 2Q^2\right)} \\
& \times \cos\left[ \frac{1}{3} \cos^{-1} \left( 
\frac{3\sqrt{3}M \left(M^2 - \alpha^2 - Q^2\right)}{ \left(3M^2 - \alpha^2 - 2Q^2\right)^{3/2}} 
\right) \right].
\end{split}
\end{equation}
Fig. \ref{PoleRPH} illustrates the variation of the photon sphere radius as a function of charge and rotation parameter. We observe that the radius of the photon sphere decreases with increasing charge as well as with increasing rotation. For comparison, we also include the cases of the SBH, KBH, and RNBH. It is evident that at the polar plane the SBH possesses the largest photon sphere radius, while the KNBH with higher charge and higher rotation exhibits the smallest photon sphere radius.
\subsection{\textbf{Equatorial $(\theta=\pi/2)$}}
In this case, we have to deal with a quartic equation to obtain the photon sphere radius \cite{Feng:2019zzn},
\begin{align} 
r_{\text{ph}}^4 
& - 6M r_{\text{ph}}^3 
+ (9M^2 + 4Q^2) r_{\text{ph}}^2 \nonumber \\
& - 4M(\alpha^2 + 3Q^2) r_{\text{ph}} 
+ 4Q^2(\alpha^2 + Q^2) = 0.
\end{align}
The equation can be solved exactly, and the subtlety lies in selecting the correct root corresponding to the photon sphere i.e. the real, positive root that lies outside the event horizon. Therefore, the radii of the photon sphere are determined by solving the quartic equation under this constraint as,\\

\begin{equation}
r_{\text{ph}}^{\pm} = -\frac{3M}{2} + \mathcal{A} \pm \frac{1}{2} \sqrt{ -4\mathcal{A}^2 - 9M^2-8Q^2 + \frac{4M \alpha^2}{\mathcal{A}} }.
\label{rpheq2}
\end{equation}
The raddi $r_{ph}^+$ and $r_{ph}^-$ represent the photon sphere for prograde and retrograde motion, respectively. The detailed expression for $\mathcal{A}$ is provided in Appendix A.
\begin{figure}
	\begin{center}
    \begin{subfigure}[]
     {\includegraphics[width=0.45\textwidth]{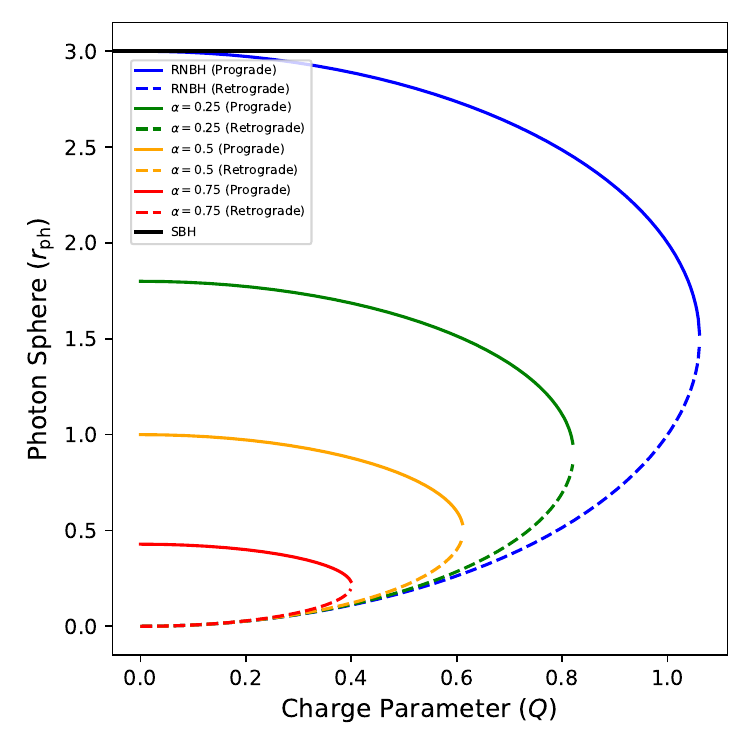}} 
    \end{subfigure}
     \begin{subfigure}[]
     {\includegraphics[width=0.45\textwidth]{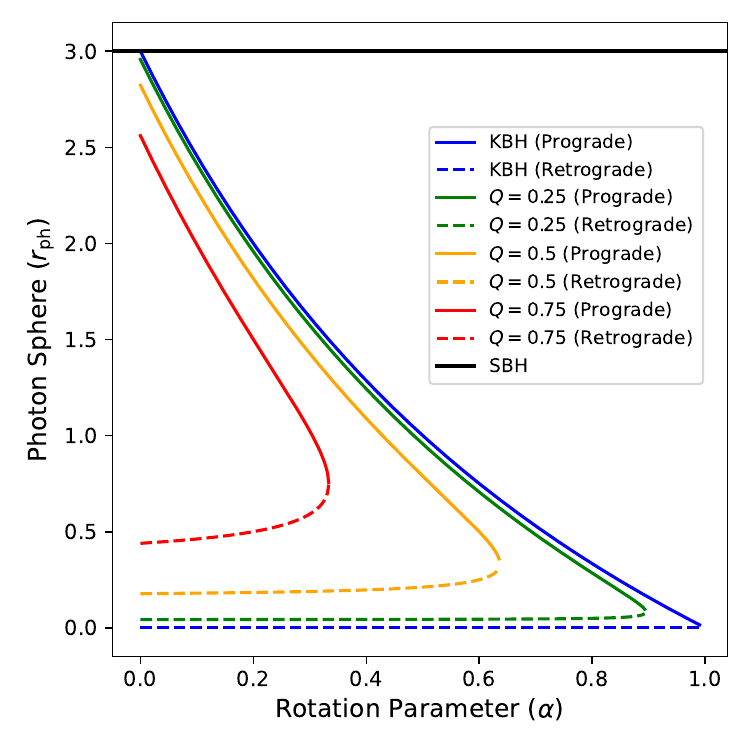}} 
    \end{subfigure}   
	\end{center}
	\caption{ The graphical representation of photon sphere radius at equatorial plane. Subfigure (a) corresponds to as a function of charge parameter with different values of rotation parameter, while subfigure (b) corresponds to as a function of rotation parameter with different values of charge parameter. Here we consider $M=1$ and $\theta=\pi/2$.} \label{EquaRPH}
\end{figure}


Fig. \ref{EquaRPH} presents the photon sphere radius as a function of the charge and rotation parameter at the equatorial plane. Solid lines represent the prograde case, while dashed lines denote the retrograde case. When plotting the photon sphere radius as a function of the charge parameter, we observe that for the prograde case, the photon sphere radius decreases with increasing charge, whereas for the retrograde case, it increases with charge. Additionally, for the maximum rotation, the photon sphere radius reaches its minimum value in the prograde case. When plotting the photon sphere radius as a function of the rotation parameter, we find that for prograde motion, the radius decreases as rotation increases. In contrast, for the retrograde case, the behavior depends on the charge: for high values of charge, the photon sphere radius increases with increasing rotation; for low charge values, the photon sphere vanishes at higher rotation. For comparison, we also include the SBH, KBH, and RNBH, to which the KNBH reduces under specific limits.

\section{Refractive Index due to Kerr-Newman Black Hole}
The linearized form of the Kerr-Newman metric (\ref{metric}) in terms of Boyer-Lindquist coordinates ($ct$, $r$, $\theta$, $\phi$) can be obtained as
\begin{align}
ds^2 =\ & \left(1 - \frac{r_g r - Q^2}{\Sigma^2} 
+ \frac{2\alpha(r_g r - Q^2)\sin^2\theta}{\Sigma^2} \frac{d\phi}{c dt} \right) c^2 dt^2 \nonumber \\
& - \frac{\Sigma^2}{\Delta} dr^2 
- \Sigma^2 d\theta^2 \nonumber \\
& - \left(r^2 + \alpha^2 + \frac{r_g r - Q^2}{\Sigma^2} \alpha^2 \sin^2\theta \right) \sin^2\theta d\phi^2.
\label{linear}
\end{align}
Under the far-field approximation $(\alpha^2/r^2 << 1)$, the above line element becomes
\begin{align}
ds^2 =\ & \left(1 - \frac{r_g}{r} + \frac{Q^2}{r^2} 
+ 2\alpha \left(\frac{r_g}{r} - \frac{Q^2}{r^2}\right) \sin^2\theta \frac{d\phi}{c dt} \right) c^2 dt^2 \nonumber \\
& - \left(1 - \frac{r_g}{r} + \frac{Q^2}{r^2} \right)^{-1} dr^2 \nonumber \\
& - r^2 \left(d\theta^2 + \sin^2\theta d\phi^2 \right).
\label{farfield}
\end{align}

Now, by coordinate transformation, it is possible to express the Kerr-Newman line element in isotropic form of the metric \cite{sen2010more, roy2015trajectory, roy2017deflection, roy2019study, roy2025deflection}. We introduced a new radial co-ordinate $\rho$ with the following transformation equation,
\begin{equation}\label{rho}
    \rho=\frac{1}{2}\left[ r - \frac{r_g}{2} +\sqrt{r^2-r r_g + Q^2}\right],
\end{equation}
or,
\begin{equation}\label{r}
   r=\rho\left[ (1+\frac{r_g}{4\rho}+\frac{Q}{2\rho})(1+\frac{r_g}{4\rho}-\frac{Q}{2\rho})\right], 
\end{equation}
which in turn leads to,
\begin{equation}\label{drdrho}
    \frac{dr}{d\rho}=1-\frac{r_g^2}{16\rho^2}+\frac{Q^2}{4\rho^2}.
\end{equation}
Now, substituting the value of $r$ and $dr^2$ from Eqs. (\ref{r}) and (\ref{drdrho}) in Eq.(\ref{farfield}) we obtain the components of a new form of the Kerr-Newman metric in coordinate system ($ct$, $\rho$, $\theta$, $\phi$) as,
\begin{align}
    g_{00} = \ & \frac{(1-\frac{r_g^2}{16\rho^2}+\frac{Q^2}{4\rho^2})^2}{\left[ (1+\frac{r_g}{4\rho})^2 - (\frac{Q}{2\rho})^2\right]^2} \nonumber \\
    & + \frac{2\alpha\left\{ \frac{r_g}{\rho} \left((1+\frac{r_g}{4\rho})^2 - (\frac{Q}{2\rho})^2\right)-\frac{Q^2}{\rho^2}
 \right\}\sin^2\theta \frac{d\phi}{c dt}}{\left[ (1+\frac{r_g}{4\rho})^2 - (\frac{Q}{2\rho})^2\right]^2}, 
\end{align}
\begin{equation}
    g_{11}=\left\{ (1+\frac{r_g}{4\rho})^2 - (\frac{Q}{2\rho})^2\right\}^2,
\end{equation}
\begin{equation}
    g_{22}=\rho^2 g_{11},
\end{equation}
\begin{equation}
    g_{33}=\rho^2 \sin^2 \theta g_{11}.
\end{equation}
Thus, the isotropic form of KNBH obtained as
\begin{strip}
\begin{align}
ds^2 = & \frac{\left(1 - \frac{r_g^2}{16\rho^2} + \frac{Q^2}{4\rho^2} \right)^2 
+ 2\alpha \left\{ \frac{r_g}{\rho} \left( \left(1 + \frac{r_g}{4\rho} \right)^2 - \left( \frac{Q}{2\rho} \right)^2 \right)
- \frac{Q^2}{\rho^2} \right\} \sin^2\theta \frac{d\phi}{c dt}}
{\left[ \left(1 + \frac{r_g}{4\rho} \right)^2 - \left( \frac{Q}{2\rho} \right)^2 \right]^2} c^2 dt^2 
  - \left\{ \left(1 + \frac{r_g}{4\rho} \right)^2  - \left( \frac{Q}{2\rho} \right)^2 \right\}^2  \nonumber \\
& \times \left\{ d\rho^2 + \rho^2 (d\theta^2 + \sin^2\theta d\phi^2) \right\}.
\label{isotropic}
\end{align}
The refractive index of the effective medium can be obtained from the isotropic form of the solution (\ref{isotropic}) by
using the material medium approach. Setting ds = 0,
the velocity of light ($v(\rho, \theta, \alpha, Q)$) can be identified from the
expression of Eq.(\ref{isotropic}) as
    \begin{equation}\label{vrho}
    v(\rho, \theta, \alpha, Q)= \frac{\sqrt{(1-\frac{r_g^2}{16\rho^2}+\frac{Q^2}{4\rho^2})^2+2\alpha\left\{ \frac{r_g}{\rho} \left((1+\frac{r_g}{4\rho})^2 - (\frac{Q}{2\rho})^2\right)-\frac{Q^2}{\rho^2}
 \right\}\sin^2\theta \frac{d\phi}{c dt}}}{\left[ (1+\frac{r_g}{4\rho})^2 - (\frac{Q}{2\rho})^2\right]^2}.
\end{equation}
However, this expression of the velocity of light is in the unit of length $\rho$ per unit of time. So, in order to express the velocity in terms of $r$ per unit time, we use Eqs. (\ref{drdrho}) and (\ref{vrho}) and write $v(r, \theta, \alpha, Q)$ as
\begin{equation}
    \begin{split}
        v(r, \theta, \alpha, Q)=& v(\rho, \theta, \alpha, Q) \frac{dr}{d\rho}\\
        =& \frac{\sqrt{(1-\frac{r_g^2}{16\rho^2}+\frac{Q^2}{4\rho^2})^2+2\alpha\left\{ \frac{r_g}{\rho} \left((1+\frac{r_g}{4\rho})^2 - (\frac{Q}{2\rho})^2\right)-\frac{Q^2}{\rho^2}
 \right\}\sin^2\theta \frac{d\phi}{c dt}}}{\left[ (1+\frac{r_g}{4\rho})^2 - (\frac{Q}{2\rho})^2\right]^2} 
 \left\{1-\frac{r_g^2}{16\rho^2}+\frac{Q^2}{4\rho^2}\right\} c\\
 =& \left\{\frac{(4\rho)^2-(r_g)^2+(2Q)^2}{(4\rho+r_g)^2-(2Q)^2}\right\}^2 \sqrt{1+32 \alpha \rho r_g \frac{(4\rho+r_g)^2-(2Q)^2-(4Q)^2}{\left\{(4\rho)^2-(r_g)^2+(2Q)^2\right\}^2}\sin^2\theta \frac{d\phi}{cdt}} c.
    \end{split}
\end{equation}
With value of $4\rho$ from Eq.(\ref{rho}), the velocity $v(r, \theta, \alpha, Q)$ becomes, 
\begin{align}
v(r, \theta, \alpha, Q) =\ & \left\{ 1 - \frac{r_g}{r} + \frac{Q^2}{r^2} \right\} \sqrt{1 + \frac{2(r_g r - Q^2)}{r^2 - r_g r + Q^2} \alpha \sin^2\theta \frac{d\phi}{c dt}}\, c.
\label{Eq_vel}
\end{align}
Thus the refractive index of the effective medium on non-equatorial plane is
\begin{align}
n(r, \theta, \alpha, Q) =\ & \left\{ 1 - \frac{r_g}{r} + \frac{Q^2}{r^2} \right\}^{-1} \left[ 1 + \frac{2(r_g r - Q^2)}{r^2 - r_g r + Q^2} \alpha \sin^2\theta \frac{d\phi}{c dt} \right]^{-1/2}.
\label{Eq_ref_r}
\end{align}
Now, by replacing $\frac{r}{r_g}$ by $x$ and $\frac{Q}{r_g}$ by $q$, we can write the above equation of the refractive index $ n(r, \theta, \alpha, Q)$ as follows,
\begin{equation} \label{Eq_ref_x}
\begin{split}
    n(r, \theta, \alpha, Q)=& \left\{ 1-\frac{1}{x} + \frac{q^2}{x^2}\right\}^{-1} \left[1+ \frac{2(x-q^2)}{x^2 - x + q^2}\alpha  \sin^2\theta \frac{d\phi}{cdt}\right]^{    -1/2}\\
    =& (1-\frac{1}{x})^{-1}\left\{1  + \frac{q^2}{x(x-1)}\right\}^{-1} \left[1+ \frac{2(x-q^2)}{x^2 - x + q^2}\alpha  \sin^2\theta \frac{d\phi}{cdt}\right]^{    -1/2}\\
    =& \left(\frac{x}{x-1}\right)\left\{1  + \frac{q^2}{x(x-1)}\right\}^{-1} \left[1+ \frac{2(x-q^2)}{x^2 - x + q^2}\alpha  \sin^2\theta \frac{d\phi}{cdt}\right]^{    -1/2}.
\end{split}
\end{equation}
\end{strip}
Setting $\alpha=0$ and $Q=0$, we find that the central gravitational mass is static and uncharged, and in such case the above expression of the refractive index goes over to that of Schwarzschild mass which is the same as the refractive index calculated by Sen\cite{sen2010more} for a static non-rotating mass (Schwarzschild geometry). Thus, the second term is in parentheses and the third term is inside the square bracket of the RHS of Eq. (\ref{Eq_ref_x}) is due to the rotation of the charged gravitating body.
The second term consists of the charge parameter $q$ and the third term consists of the rotation parameter $\alpha$ and the frame-dragging parameter $\frac{d\phi}{cdt}$ due to the rotating body. For a static uncharged body, these terms just vanish. However, for a charged and very slow rotating body, the RHS of Eq. (\ref{Eq_ref_x}) can be expanded as an infinite converging series, since the second and third terms are significantly less than 1. It may be noted here that Sen’s\cite{sen2010more} work is restricted to static uncharged geometry. But Eq. (\ref{Eq_ref_x}) can include the case of a charged rotating body (stationary geometry). Therefore, the case discussed by Sen\cite{sen2010more}, appears as a special case when $\alpha$ and $Q$ are set to zero.
\begin{figure*}
	\begin{center}
    \begin{subfigure}[]
     {\includegraphics[width=0.45\textwidth]{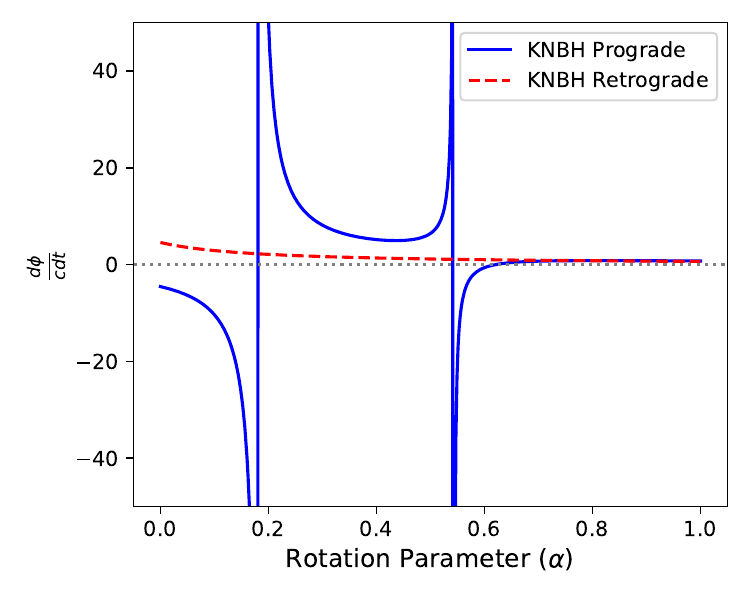}}   
    \end{subfigure}
     \begin{subfigure}[]
     {\includegraphics[width=0.45\textwidth]{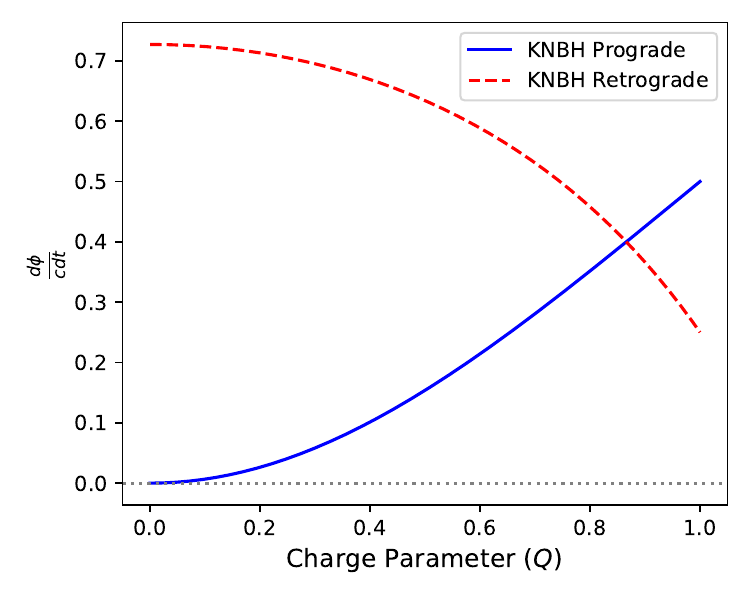}} 
    \end{subfigure}   
    \begin{subfigure}[]
     {\includegraphics[width=0.45\textwidth]{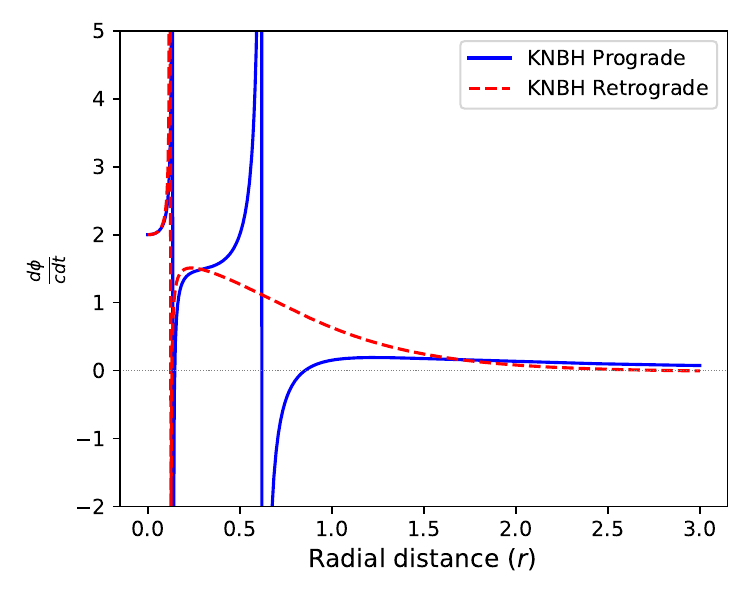}} 
    \end{subfigure} 
    \begin{subfigure}[]
     {\includegraphics[width=0.45\textwidth]{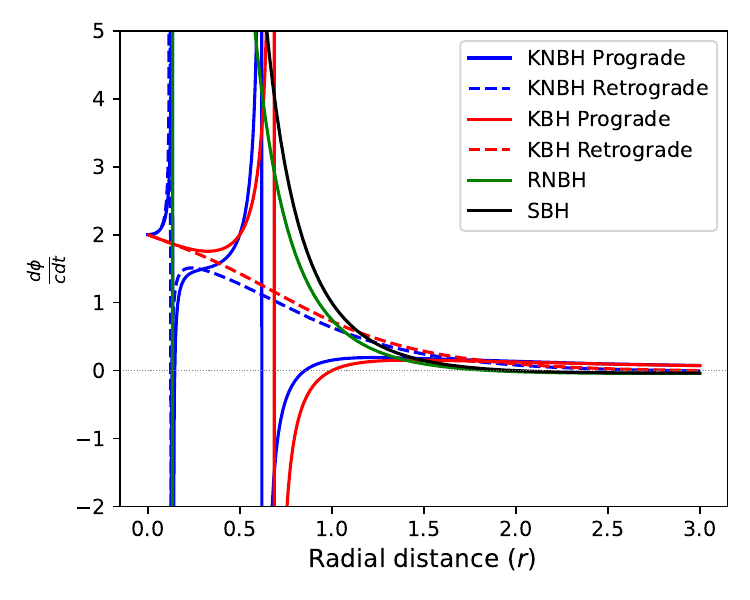}} 
    \end{subfigure} 
    \begin{subfigure}[]
     {\includegraphics[width=0.45\textwidth]{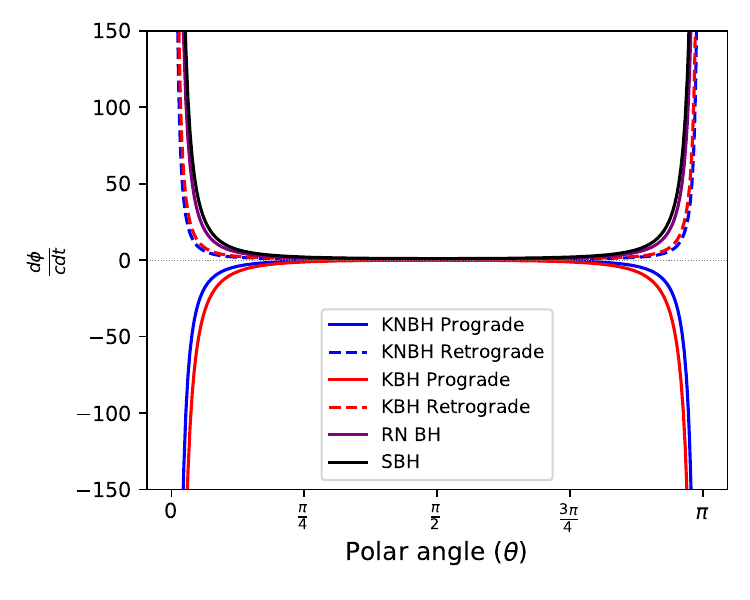}} 
    \end{subfigure}
    \begin{subfigure}[]
     {\includegraphics[width=0.45\textwidth]{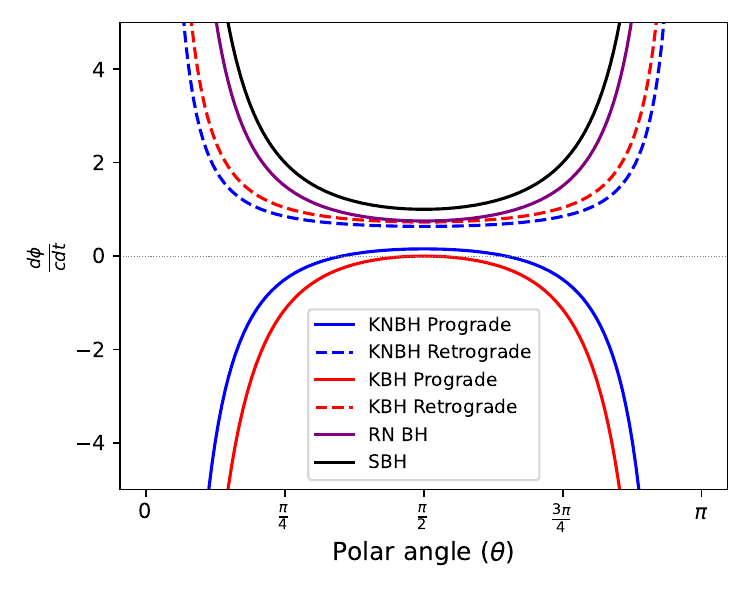}} 
    \end{subfigure}
	\end{center}
	\caption{The variation of frame dragging as a function of the rotation parameter, charge parameter, radial distance, and polar angle, along with a comparison to other BHs solution in GR. Subfigure (f) is the zoom in view of subfigure (e). Here we consider $M=1$ and $L=\mp 0.5 (Pro/Retro)$. We used $q=0.3$, $u=0.4$, $\theta=\pi/2$ and $r$ near the horizon, where frame dragging not plotted as a function of these parameters.} \label{FDfig}
\end{figure*}
\subsection{\textbf{The Frame-dragging}}
The frame-dragging effect due to the Kerr-Newman space-time can be obtained by using the four-momentum equation \cite{roy2015trajectory, roy2025deflection}, we have the following. 

The four-momentum of the particle is
 \begin{equation}
     p^i = mc \frac{dx^i}{d\lambda}= g^{ik}P_k = -g^{ik} \frac{\partial S}{\partial x^k} ,
\end{equation}
 where $i$ and $k$ have the values 0, 1, 2, and 3 which stand for
the coordinates $ct$, $r$, $\theta$, $\phi$ respectively, $\lambda$ is the affine parameter and $S = - E_0 t  + L \phi +S_r (r) + S_\theta (\theta)$ is the relativistic action function for a particle in the gravitational field of a rotating spherical mass, with $E_0$ as the conserved energy and $L$ denotes the component of the angular momentum along the axis of the symmetry of the field. Now, for the variables $t$ and $\phi$, we can write: 

\begin{equation}\label{Eq_ct}
    mc \frac{dx^0}{d\lambda} = -g^{00} \frac{\partial S}{\partial x^0} -g^{01} \frac{\partial S}{\partial x^1} - g^{02} \frac{\partial S}{\partial x^2} - g^{03} \frac{\partial S}{\partial x^3}
\end{equation} 

and 
\begin{equation}\label{Eq_phi}
     mc \frac{dx^3}{d\lambda} = -g^{30} \frac{\partial S}{\partial x^0} -g^{31} \frac{\partial S}{\partial x^1} - g^{32} \frac{\partial S}{\partial x^2} - g^{33} \frac{\partial S}{\partial x^3}.
\end{equation} 

Compared with the Kerr-Newman line element expressed by Eq.(\ref{metric}), the required contra-variant components are

 \begin{equation}\label{g_co_comp}
     \begin{split}
         g^{00} = &\frac{1}{\Delta} \left(r^2+\alpha^2+\frac{r_gr-Q^2}{\Sigma^2}\alpha^2 \sin^2\theta\right),\\
         g^{03}=& g^{30}= \frac{r_gr-Q^2}{\Delta \Sigma^2}\alpha,\\
         g^{33}= & - \frac{\Sigma^2 - (r_gr-Q^2)}{\Delta \Sigma^2 \sin^2\theta},\\        
     \end{split}
 \end{equation}
and other components are zero.
 
Using the above values of the components, expressed in Eq.(\ref{g_co_comp})  of the KNBH, we have generated the Eqs.(\ref{Eq_ct}) and (\ref{Eq_phi}) as
\begin{align}
    mc^2 \frac{dt}{d\lambda}=&-\frac{1}{\Delta} \left(r^2+\alpha^2+\frac{r_gr-Q^2}{\Sigma^2}\alpha^2\sin^2\theta\right)\left(-\frac{E_0}{c}\right) \nonumber\\
    &- \left(\frac{r_gr-Q^2}{\Delta \Sigma^2}\alpha\right)L ,
\end{align}
 and, 
 \begin{equation}
     mc\frac{d\phi}{d\lambda}= -\left(\frac{r_gr-Q^2}{\Delta \Sigma^2}\alpha\right)\left(-\frac{E_0}{c}\right) + \frac{\Sigma^2 - (r_gr-Q^2)}{\Delta \Sigma^2 \sin^2\theta}L.
 \end{equation} 
\begin{strip}
Thus, the frame-dragging of KNBH  is obtained as
    \begin{equation} \label{Eq_FD_G}
\begin{split}
     \frac{d\phi}{cdt} = & \frac{\left[(r_gr-Q^2)\alpha\sin^2\theta \frac{E_0}{c}\right] + \left[\Sigma^2-(r_gr-Q^2)\right]L}{\left[\Sigma^2\left(r^2+\alpha^2+\frac{r_gr-Q^2}{\Sigma^2}\alpha^2\sin^2\theta\right) \frac{E_0}{c} - (r_gr-Q^2)\alpha L\right]\sin^2\theta}\\
    =&  \frac{\left[(r_gr-Q^2)\alpha\sin^2\theta \right] + \left[\Sigma^2-(r_gr-Q^2)\right]b }{\left[\Sigma^2\left(r^2+\alpha^2+\frac{r_gr-Q^2}{\Sigma^2}\alpha^2\sin^2\theta\right)  - (r_gr-Q^2)\alpha b \right]\sin^2\theta}.
\end{split}
 \end{equation}

This is the general expression of the frame dragging of KNBH in the non-equatorial plane in a strong field limit. 
Here, it should be noted that in the propagation of a light ray (or a photon-like particle) parallel to the x-axis, i.e. $\phi$ is restricted to $\pm \pi/2$ (at the location where the impact parameter is measured), the momentum $(p)$ and the conserved energy $(E_0)$ can be expressed by relation $E_0=pc$ and the angular momentum $L$ as
$L=pb$, where impact parameter $=\sqrt{y^2+z^2}=r \sqrt{\sin^2 \theta \sin^2 \phi+\cos^2 \theta}=r\sqrt{\sin^2 \theta +\cos^2 \theta}=r=b$ ($b$ is the physical radius of KNBH). 


Now, under the far-field approximation the frame dragging of KNBH in the non-equatorial plane becomes,
\begin{equation} \label{Eq_FD_far}
     \frac{d\phi}{cdt} = \frac{\left[(r_gr-Q^2)\alpha\sin^2\theta \right] + \left[r^2-(r_gr-Q^2)\right]b }{\left[r^4 - (r_gr-Q^2)\alpha b \right]\sin^2\theta}.
 \end{equation}

Furthermore, if we set $\theta=\frac{\pi}{2}$ in Eq.(\ref{Eq_FD_far}), the frame dragging for KNBH can be obtained in the equatorial plane as,
\begin{equation} \label{Eq_FD_far_eqa}
     \frac{d\phi}{cdt} = \frac{\left[(r_gr-Q^2)\alpha\right] + \left[r^2-(r_gr-Q^2)\right]b }{\left[r^4 - (r_gr-Q^2)\alpha b \right]}.
 \end{equation}
By substituting the value of $\frac{d\phi}{cdt}$ from Eq.(\ref{Eq_FD_far}) into Eq.(\ref{Eq_ref_r}) or Eq. (\ref{Eq_ref_x})
the refractive index at an arbitrary point on the non-equatorial plane in the KNBH field is
\begin{equation}\label{Eq_ref_r_FD}
\begin{split}
    n(r, \theta, \alpha, Q)= &\left\{ 1-\frac{r_g}{r} + \frac{Q^2}{r^2}\right\}^{-1} \left[1+ \frac{2(r_gr-Q^2)\alpha}{r^2 - r_g r + Q^2 } \times    \frac{\left[(r_gr-Q^2)\alpha\sin^2\theta \right] + \left[r^2-(r_gr-Q^2)\right]b }{\left[r^4 - (r_gr-Q^2)\alpha b \right]}\right]^{-1/2},
\end{split}
\end{equation}
or,
\begin{equation} \label{Eq_ref_x_FD}
\begin{split}
    n(x, \theta, \alpha, Q)=& \left(\frac{x}{x-1}\right)\left\{1  + \frac{q^2}{x(x-1)}\right\}^{-1} \left[1+ \frac{2(x-q^2)u}{x^2 - x + q^2}  \times \frac{\left[(x -q^2)u\sin^2\theta \right] + \left[x^2-(x-q^2)\right]v }{\left[x^4 - (x-q^2)u v \right]}\right] ^{-1/2}\\
    =&  n_0(x) (1+C_x)^{-1} (1+ 2S_x) ^{-\frac{1}{2}},
\end{split}
\end{equation}
\end{strip}
where we have introduced some notations as $\frac{r}{r_g}=x$, $\frac{Q}{r_g}=q$, $\frac{\alpha}{r_g}=u$, $\frac{b}{r_g}=v$ and defined 
\begin{equation}
\label{eq:n0_Cx_Sx}
\begin{split}
    n_0 =& \left( \frac{x}{x - 1} \right), \\
    C_x =& \frac{q^2}{x(x - 1)}, \\
    S_x =& \frac{(x - q^2)u}{x^2 - x + q^2} 
    \times \frac{(x - q^2) u \sin^2 \theta + [x^2 - (x - q^2)] v } {[x^4 - (x - q^2) u v] } .
\end{split}
\end{equation}
The Eq.(\ref{Eq_ref_r_FD}) or (\ref{Eq_ref_x_FD}) is the general expression of the refractive index on the non-equatorial plane in the field of rotating charged body.
Now, it is possible to show that for all $r>>r_g$ and $r>>Q$, we must have $x>>1$ and $x>>q$. As $x>>1$, we can approximate $(x-1)\approx x$ and then we can finally show that $C_x<<1$. There is another possibility that, when $r >> r_g$ and $r >> \alpha$, we must have $x >> 1$ and $x >> u$. Also, since $\alpha<b$, we must have $u<v$ and finally $S_x <<1$. So, from the assumption of $r>>r_g$, $r>>Q$ and $r >> \alpha$ that results $ C_x << 1$ and $S_x <<1$, we can say $ n <  n_0$ i.e., the refractive index in the field of rotating charged body is less than in the field of static body. If we set $\alpha=0$, the gravitational mass becomes charged and the above Eq.(\ref{Eq_ref_r_FD}) or (\ref{Eq_ref_x_FD}) goes to that for RN geometry \cite{roy2017deflection}. If we set only the charge parameter $(q)$ as zero, then the refractive index in the equatorial plane ($\theta=\frac{\pi}{2}$) becomes that of the Kerr geometry \cite{roy2015trajectory}. But if both charge and rotation parameters vanish, the central gravitational mass is static and uncharged and in that case, the above expression of the refractive index goes over to that of Schwarzschild \cite{sen2010more}. 
\begin{figure} 
	\begin{center}
    \begin{subfigure}[]
     {\includegraphics[width=0.35\textwidth]{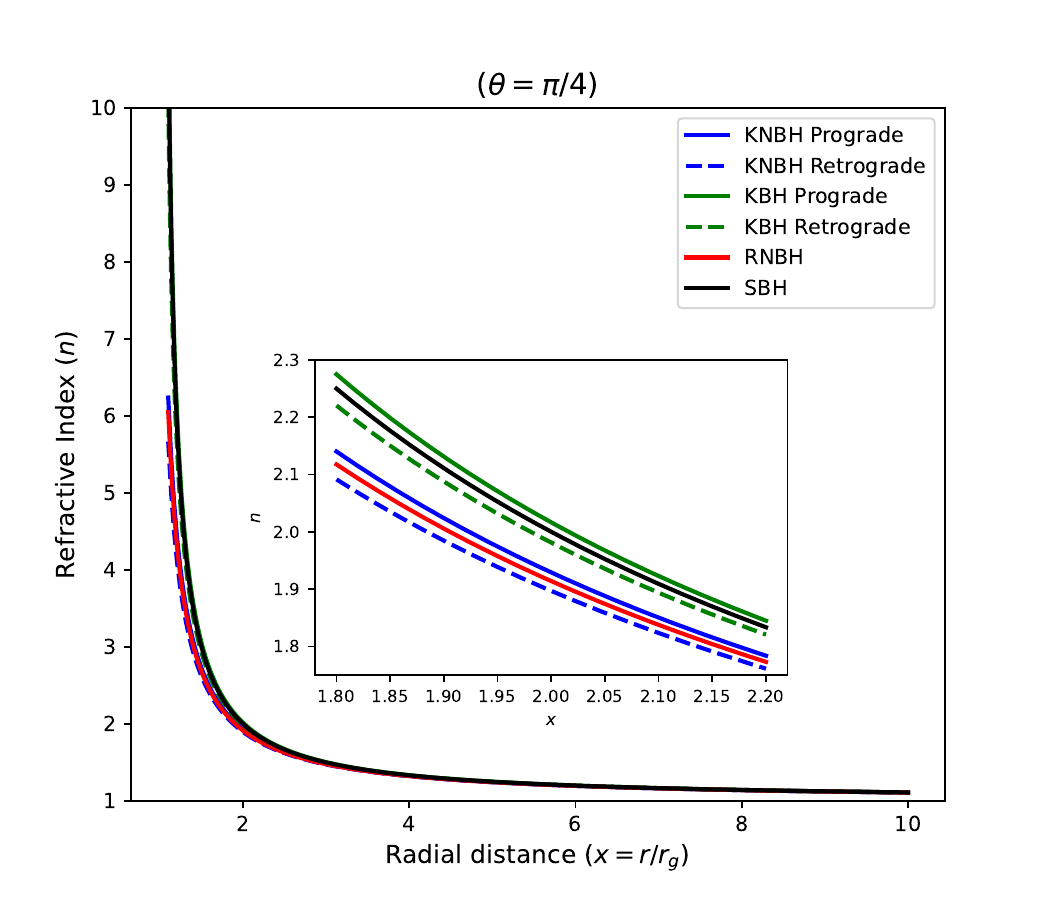}}   
    \end{subfigure}
     \begin{subfigure}[]
     {\includegraphics[width=0.35\textwidth]{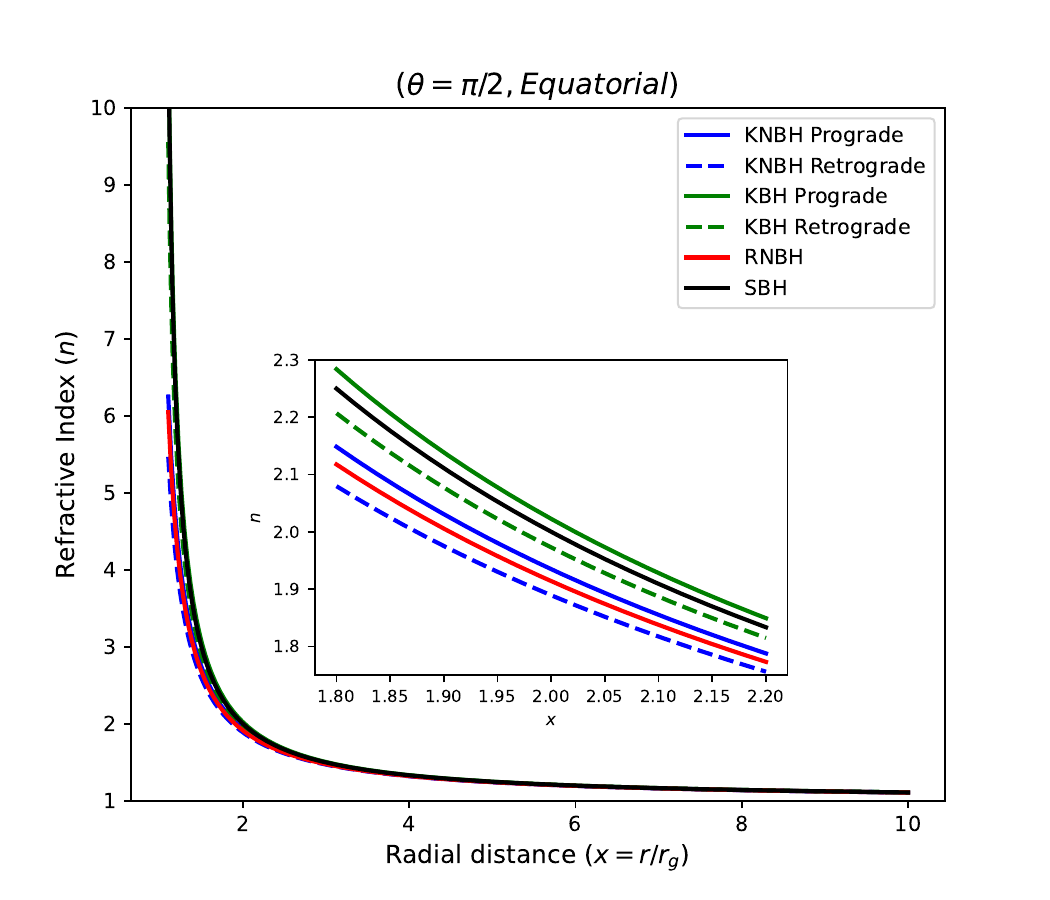}} 
    \end{subfigure}  
    \begin{subfigure}[]
     {\includegraphics[width=0.35\textwidth]{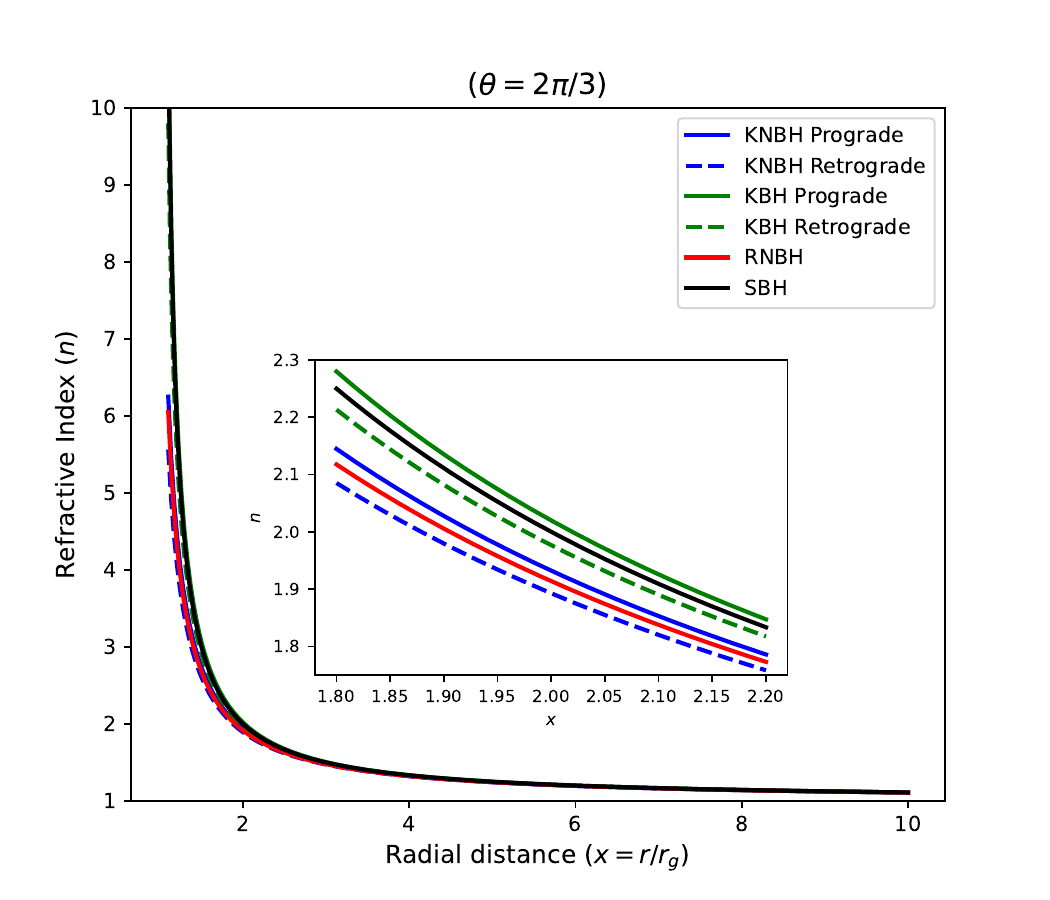}}   
    \end{subfigure}
     \begin{subfigure}[]
     {\includegraphics[width=0.35\textwidth]{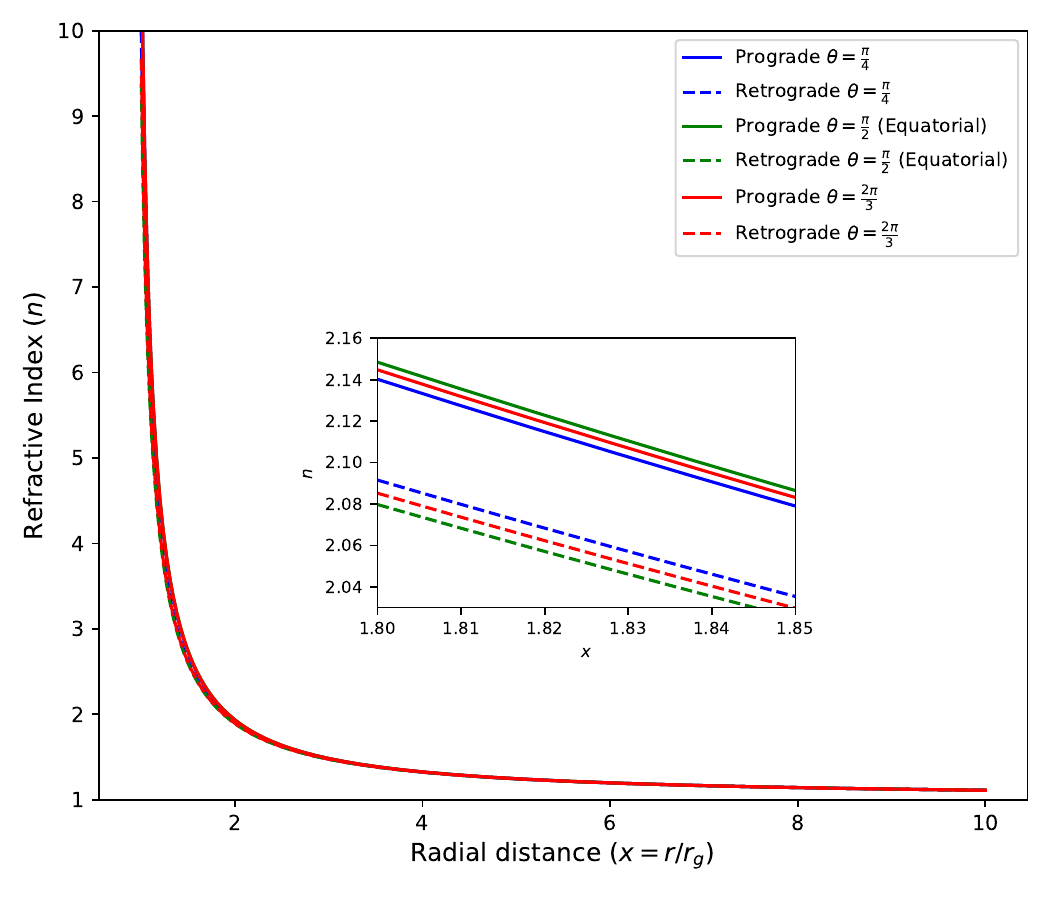}} 
    \end{subfigure}  
	\end{center}
	\caption{The variation of refractive index with radial distance. Here we consider $M=1$, $u=0.4$, $q=0.5$ and $v=\pm4(Retro/Pro)$.} \label{RIRadial}
\end{figure}
\begin{figure} 
	\begin{center}
    \begin{subfigure}[]
     {\includegraphics[width=0.35\textwidth]{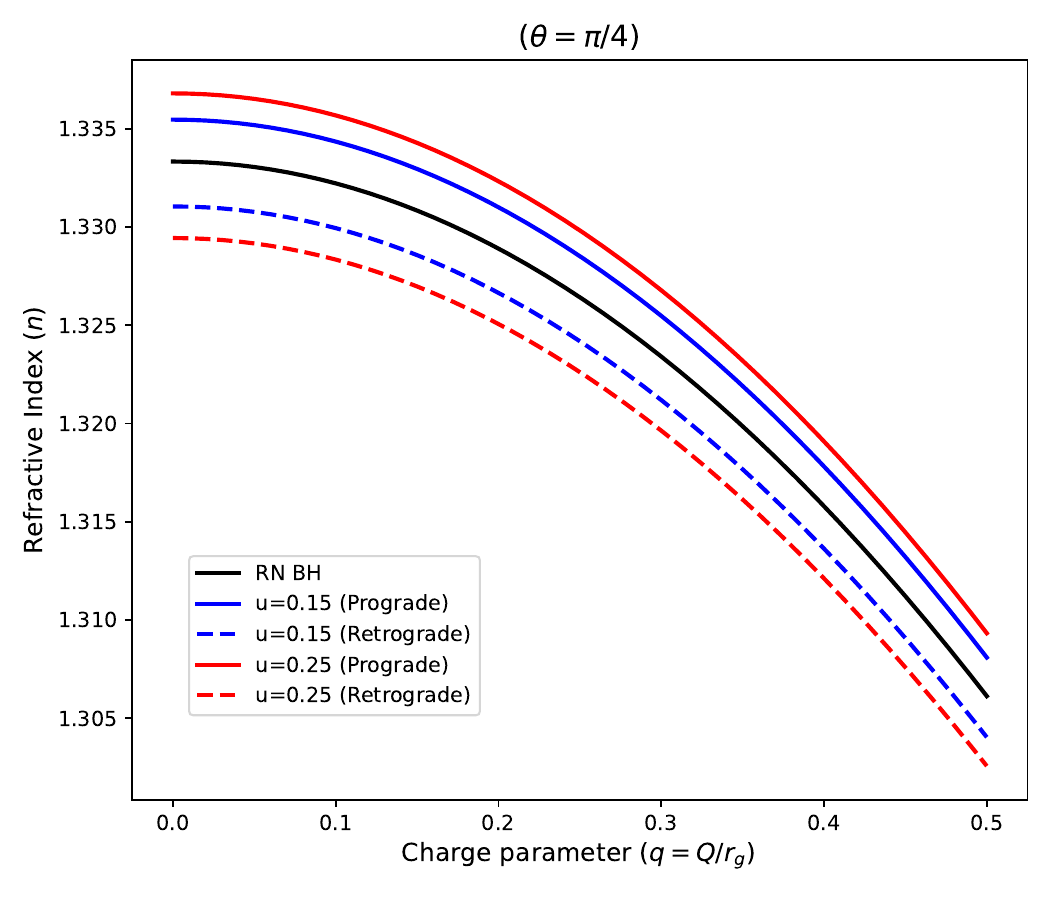}}   
    \end{subfigure}
     \begin{subfigure}[]
     {\includegraphics[width=0.35\textwidth]{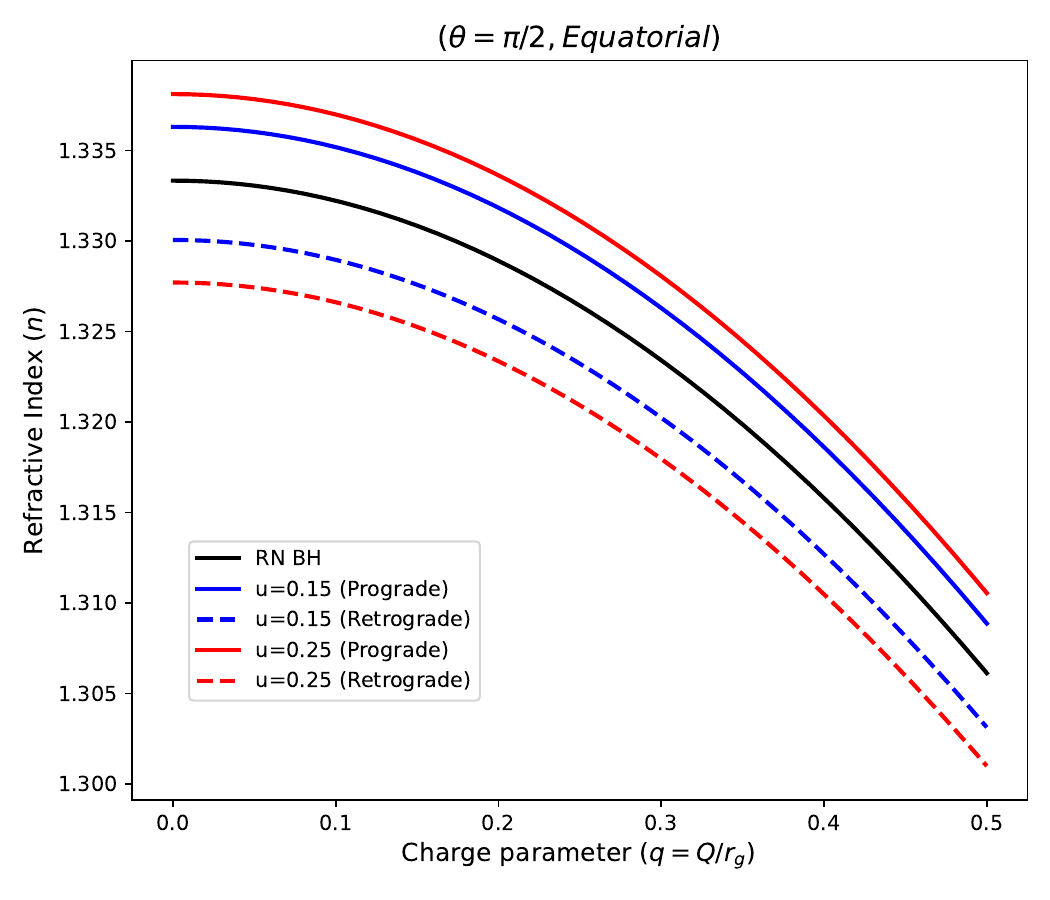}} 
    \end{subfigure}  
    \begin{subfigure}[]
     {\includegraphics[width=0.35\textwidth]{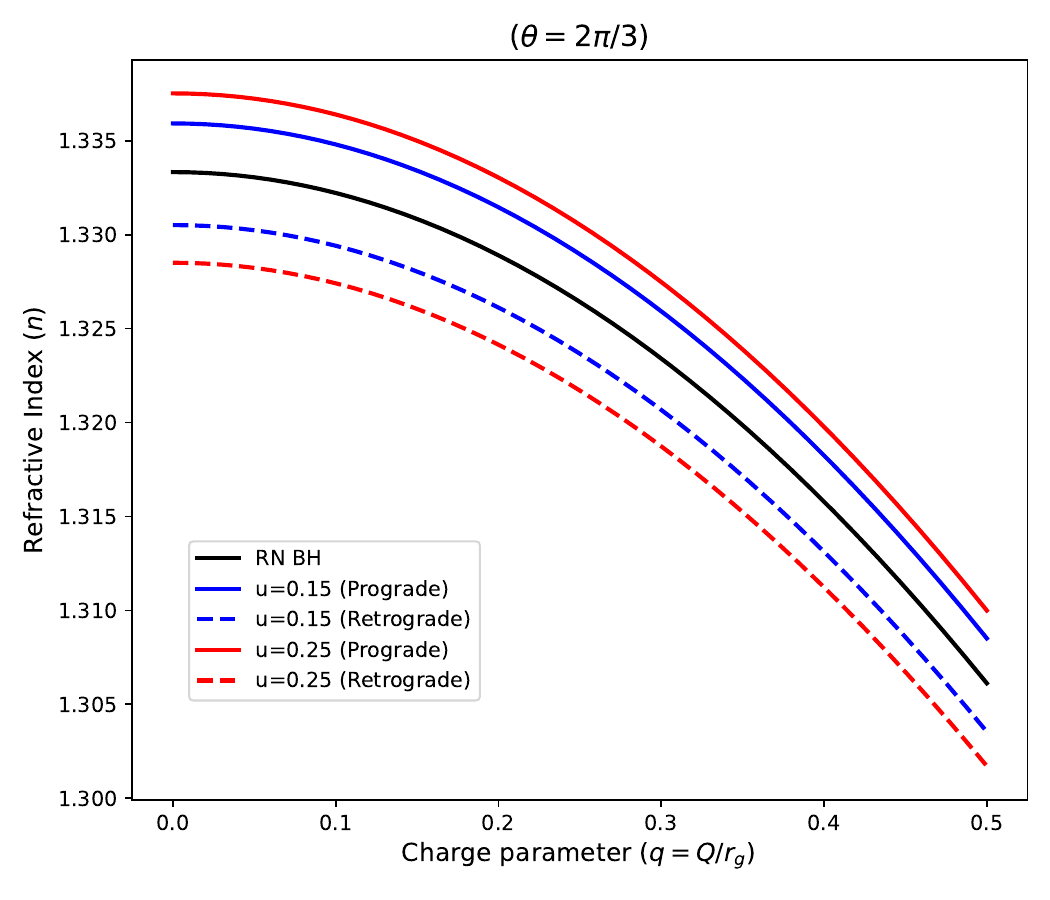}}   
    \end{subfigure}
     \begin{subfigure}[]
     {\includegraphics[width=0.35\textwidth]{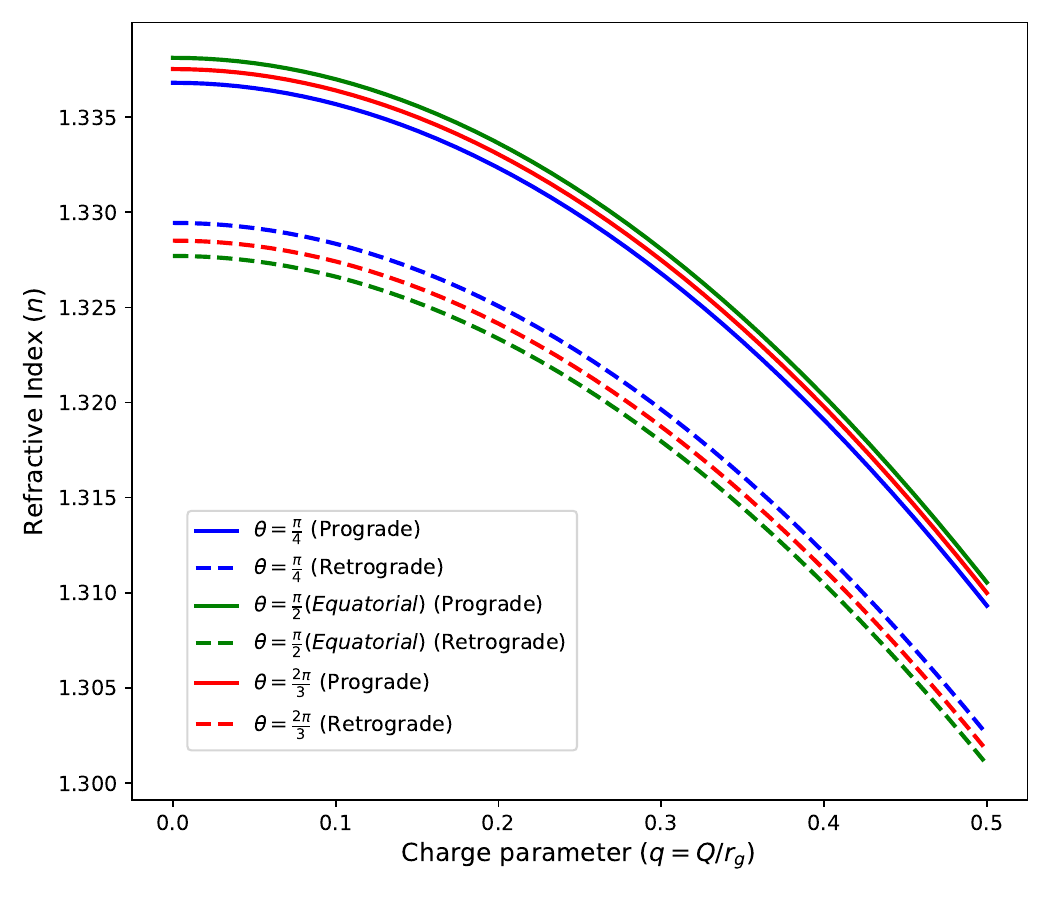}} 
    \end{subfigure}  
	\end{center}	\caption{The variation of refractive index with charge parameter. Here we consider $M=1$ and $x=10$ (subfig (d)).}  \label{RIQ}
\end{figure}
\begin{figure} 
	\begin{center}
    \begin{subfigure}[]
     {\includegraphics[width=0.35\textwidth]{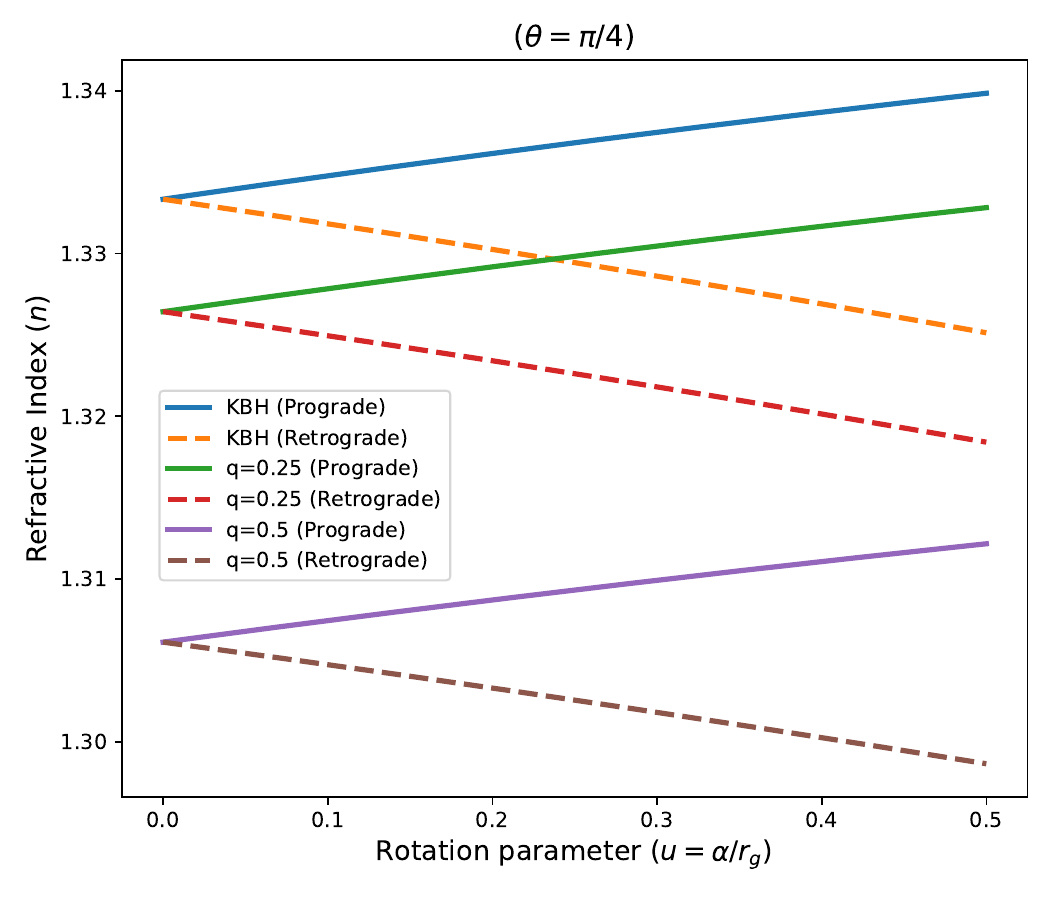}}   
    \end{subfigure}
     \begin{subfigure}[]
     {\includegraphics[width=0.35\textwidth]{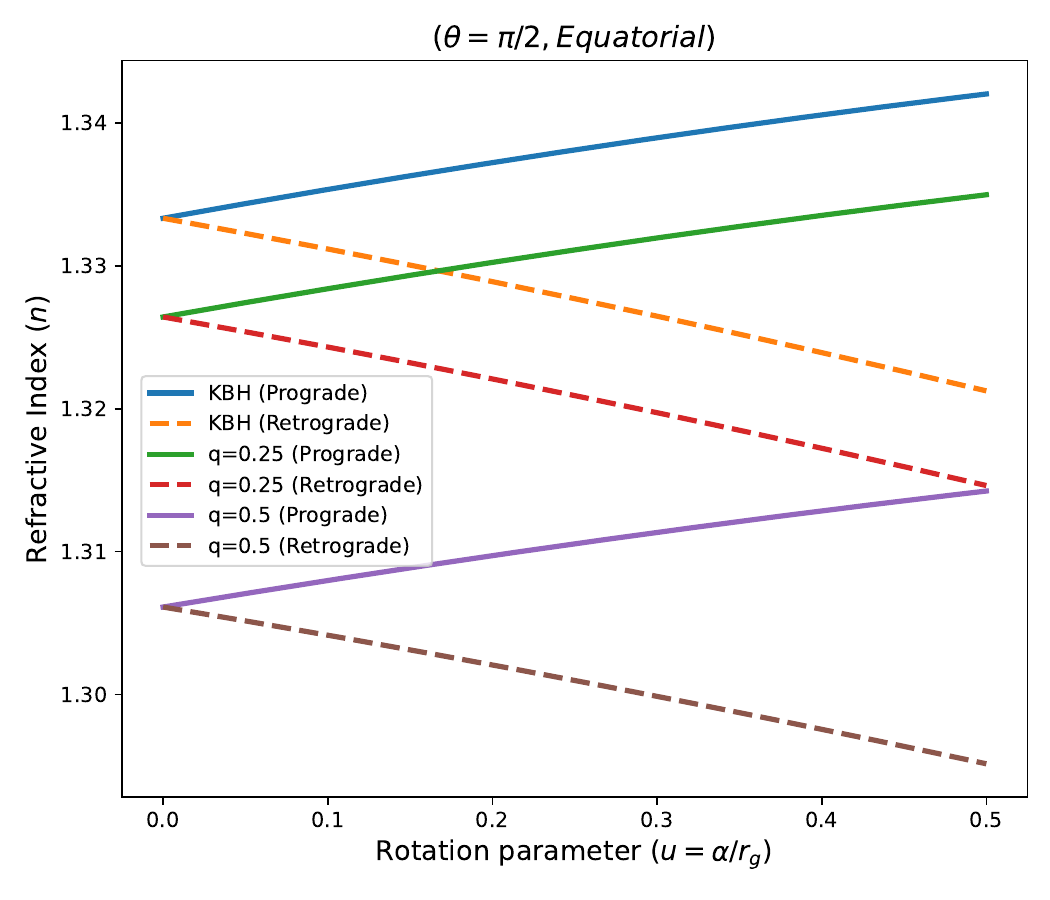}} 
    \end{subfigure}  
    \begin{subfigure}[]
     {\includegraphics[width=0.35\textwidth]{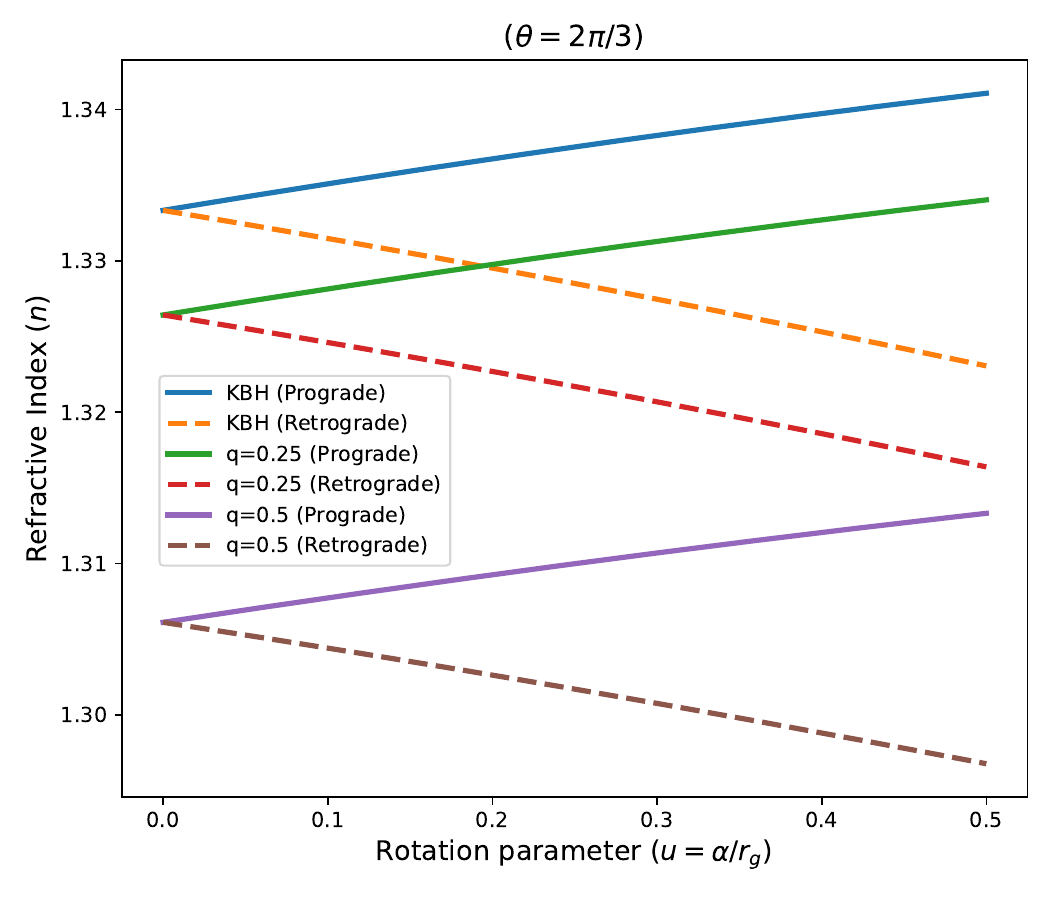}}   
    \end{subfigure}
     \begin{subfigure}[]
     {\includegraphics[width=0.35\textwidth]{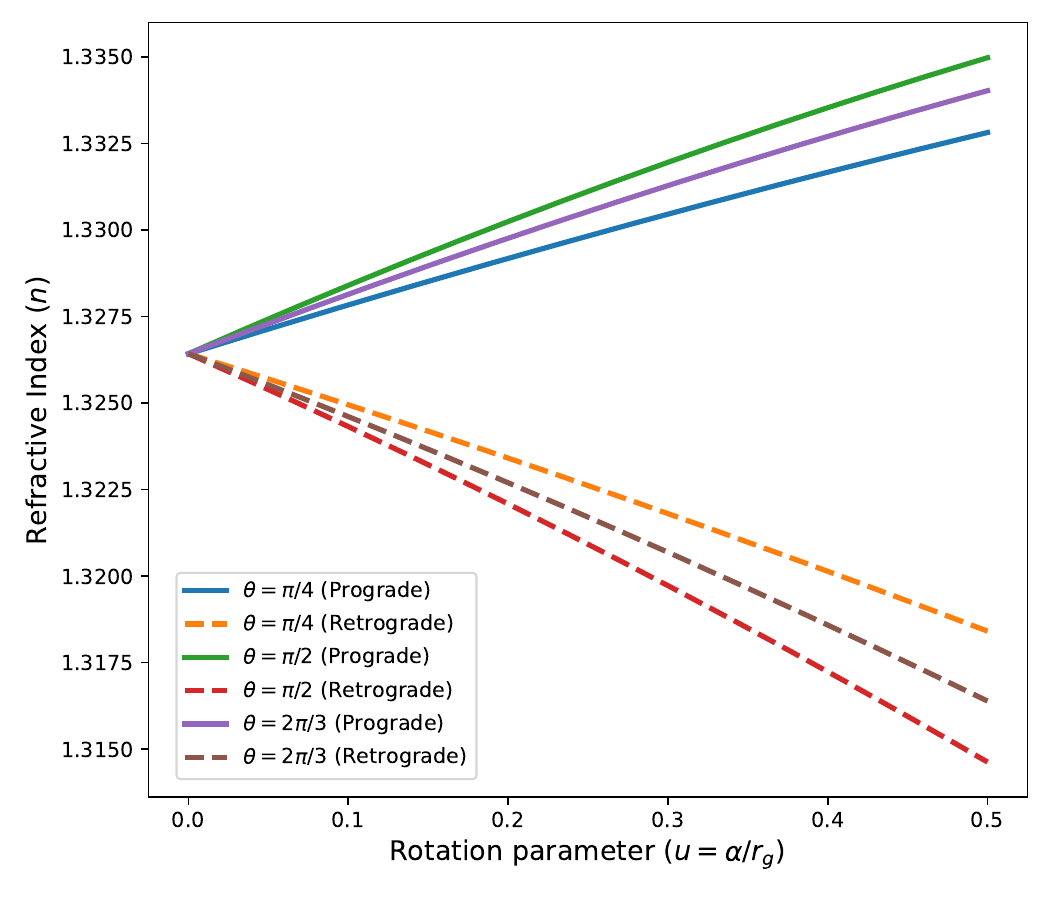}} 
    \end{subfigure}  
	\end{center}
	\caption{The variation of refractive index with rotation parameter. Here we consider $M=1$, $u=0.4$ and $x=10$ (subfig (d)).} \label{RIa}
\end{figure}

Now, by substituting $\theta=\frac{\pi}{2}$ in Eq.(\ref{Eq_ref_r_FD}) or (\ref{Eq_ref_x_FD}), the refractive index due to KNBH can be obtained in the equatorial plane as 
\begin{equation}\label{Eq_ref_r_FD_equa}
\begin{split}
    n(r, \pi/2, \alpha, Q)= \left\{ 1-\frac{r_g}{r} + \frac{Q^2}{r^2}\right\}^{-1} \\
     \times \left[1+ \frac{2(r_gr-Q^2)\alpha}{r^2 - r_g r + Q^2}   \right]^{-1/2} \\
 \times  \left[ \frac{\left[(r_gr-Q^2)\alpha \right] + \left[r^2-(r_gr-Q^2)\right]b }{\left[r^4 - (r_gr-Q^2)\alpha b \right]}\right]^{-1/2},
\end{split}
\end{equation}
or,
\begin{equation} \label{Eq_ref_x_FD_equa}
\begin{split}
    n(x, \pi/2, \alpha, Q)=& \left(\frac{x}{x-1}\right)\left\{1  + \frac{q^2}{x(x-1)}\right\}^{-1} \\
    & \times \left[1+ \frac{2(x-q^2)u}{x^2 - x + q^2}  \right]^{-1/2} \\
    & \times \left[ \frac{\left[(x -q^2)u \right] + \left[x^2-(x-q^2)\right]v }{\left[x^4 - (x-q^2)u v \right]}\right] ^{-1/2}\\
    =&  n_0(x) (1+C_x)^{-1} (1+ 2S_x^\prime) ^{-\frac{1}{2}},
\end{split}
\end{equation}
where $n_0=(\frac{x}{x-1})$  and $C_x=\frac{q^2}{x(x-1)}$ are same as before but $S_x^\prime$ for equatorial plane as
\begin{equation}
    \begin{split}
    S_x^\prime =&\left [ \frac{(x-q^2)u}{x^2-x+q^2}  \times \frac{(x-q^2)u + [x^2-(x-q^2)]v }{[x^4-(x-q^2)uv ]} \right].
    \end{split}
\end{equation}

In Fig.\ref{FDfig} the frame dragging effect of KNBH is illustrated as a function of rotation parameter($\alpha$), charge parameter($Q$), radial distances($r$), and polar angle($\theta$) respectively. In Figs. \ref{FDfig}(a) to (c) different color encodes the prograde and retrograde motion of KNBH. In Fig. \ref{FDfig}(d) and (e), the frame dragging of KNBH is illustrated and compared with other BH solutions (KBH, RNBH and SBH), where different colors encode different BHs and retrograde motions are plotted with a dashed line. As mentioned in the caption of this figure, Figs. (a) to (d) are plotted in the equatorial plane.  

Fig.\ref{FDfig}(a) is plotted at fixed rotation parameter $q=Q/r_g=0.3$. In this figure, the frame dragging effect is more prominent in prograde motion than in retrograde motion. SBH and RNBH have no dependence on the rotation parameter, and in the presence of spin on BH in the same direction as the light ray, the drag is greater than in the opposite direction of the light ray.
Fig.\ref{FDfig}(b) is plotted at fixed rotation parameter $u=\alpha/r_g =0.4$,  showing that frame dragging decreases with charge in retrograde and increases in prograde. However, at $Q=0$ and the non-zero rotation parameter, the convergent nature of the trajectories reveals the prograde and retrograde nature of the BH. At higher charge values, both trajectories coincide, resulting in the conclusion that at this point the frame dragging effect is the same for prograde and retrograde motion. Thus, depending on the values of rotation parameter and charge parameter, the frame dragging effect would be the same for both prograde and retrograde motion under KNBH at one point, and this nontrivial nature results because of the interplay between charge and rotation.  
In Figs.\ref{FDfig}(c) and (d),  the frame dragging is plotted against the radial distances with $q=0.3$ and $u=0.4$.  In the absence of the charge parameter (q=0), the frame dragging coincides with that of KBH. Similarly, in the absence of the rotation parameter ($\alpha=0$), corresponds to RNBH  and in the absence of the charge parameter and rotation parameter ($q=0$ and $\alpha=0$), the frame dragging corresponds to SBH. 
Prograde trajectories usually exhibit a higher drag than retrograde ones, and in the presence of charge, dragging is more prominent.
Figs.\ref{FDfig}(e) and (f) reveal that the dragging of BH is more significant near the pole and the anti-pole position for all BH solutions in GR. 
Another important fact to be noted that frame dragging increases towards poles for retrograde trajectories while decreasing for prograde trajectories. Thus, the magnitude of frame dragging is higher near the pole surface than the surface of the equatorial plane. This is due to the rotation frequency of the BH for which the pole is closer to the center than the equatorial surface.
All BH models are more distinctive in smaller radial distances and polar angles, indicating their independence in the BH parameters.

Figs \ref{RIRadial} - \ref{RIa} collectively present a comprehensive graphical analysis of the refractive index in the KNBH spacetime, highlighting its dependence on radial distance, charge parameter, rotation parameter, and polar angle. Fig. \ref{RIRadial} shows the variation of the refractive index as a function of the radial distance for different values of the polar angle $(\theta)$. Subfigures (a)–(c) correspond to an increasing polar angle, with the subfigure (b) specifically highlighting the equatorial plane $(\theta=\pi/2)$. In each case, the KNBH results are compared with those of well-known solutions in GR, namely the KBH, RNBH, and SBH. Subfigure (d) focuses solely on the KNBH, showing the impact of varying $(\theta)$ on the refractive index. It is evident that the refractive index is most enhanced in the KBH prograde case and least pronounced in the KNBH retrograde case, particularly in the equatorial plane. In the context of KNBH, equatorial pro is most enhanced, whereas equatorial retro is least pronounced. Fig. \ref{RIQ} illustrates the variation of the refractive index as a function of the charge parameter. The results indicate a monotonic decrease in the refractive index with increasing charge. The influence of rotation parameter is also clearly evident. In the prograde case, the refractive index increases with increasing rotation, whereas in the retrograde case, it decreases. In the limiting case of non-rotation, the KNBH reduces to the RNBH, represented by the solid black line. Consistent with the radial dependence observed previously, the refractive index attains its maximum on the equatorial plane in the prograde case and its minimum in the retrograde case. Fig. \ref{RIa} depicts the dependence of the refractive index on the rotation parameter. The results indicate that for prograde trajectories, the refractive index increases with increasing rotation, while for retrograde trajectories, it exhibits a decreasing trend. Additionally, variations in the charge parameter are examined. In both prograde and retrograde regimes, the refractive index is found to decrease with increasing charge, attaining its maximum in the uncharged limit (i.e. the KBH). Moreover, the refractive index is found to be highest on the equatorial plane for prograde motion and gradually decreases towards the polar plane.
\begin{figure}
	\begin{center}
    \begin{subfigure}[]
     {\includegraphics[width=0.5\textwidth,height=0.34\textheight]{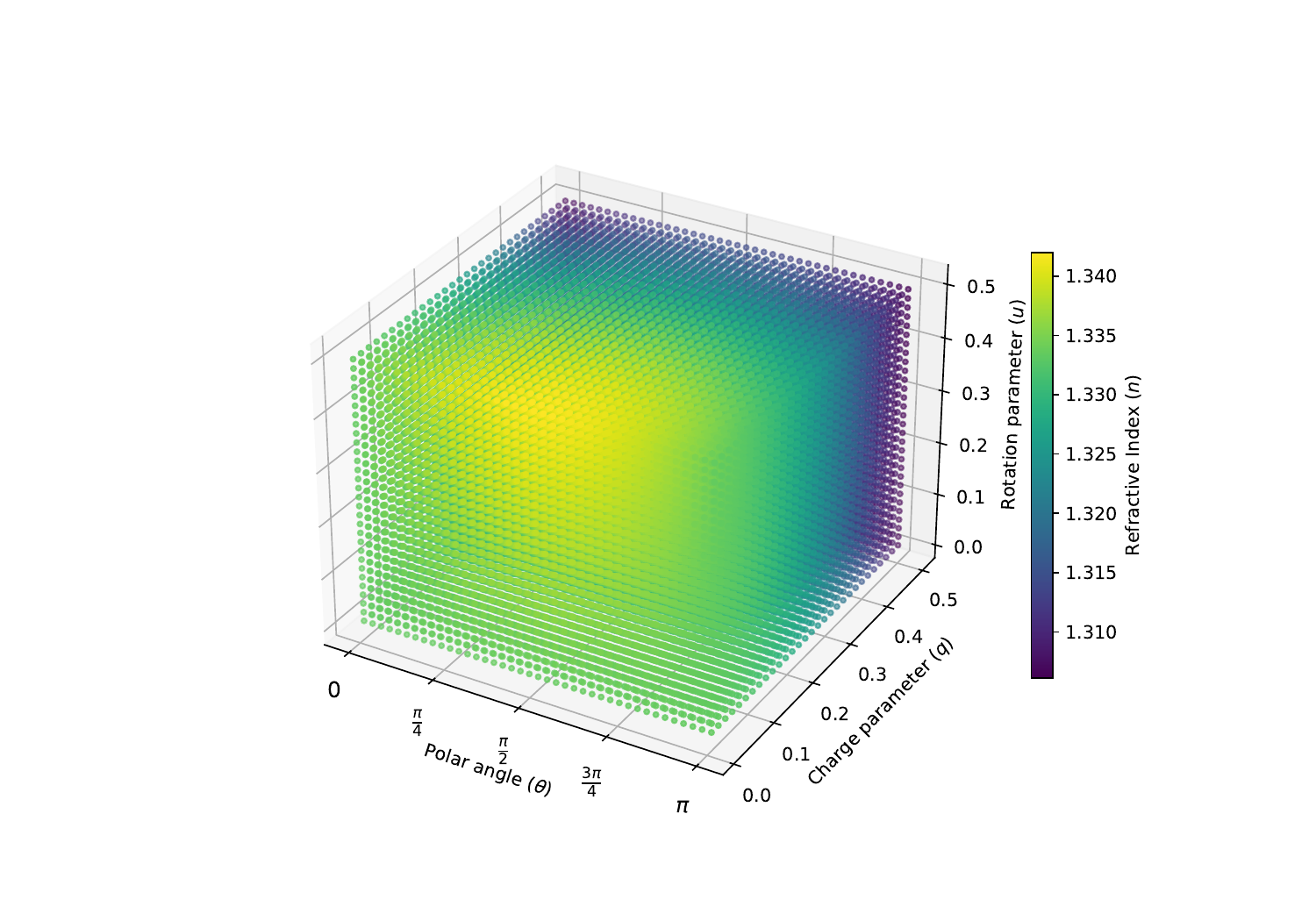}} 
    \end{subfigure}
     \begin{subfigure}[]
     {\includegraphics[width=0.5\textwidth,height=0.34\textheight]{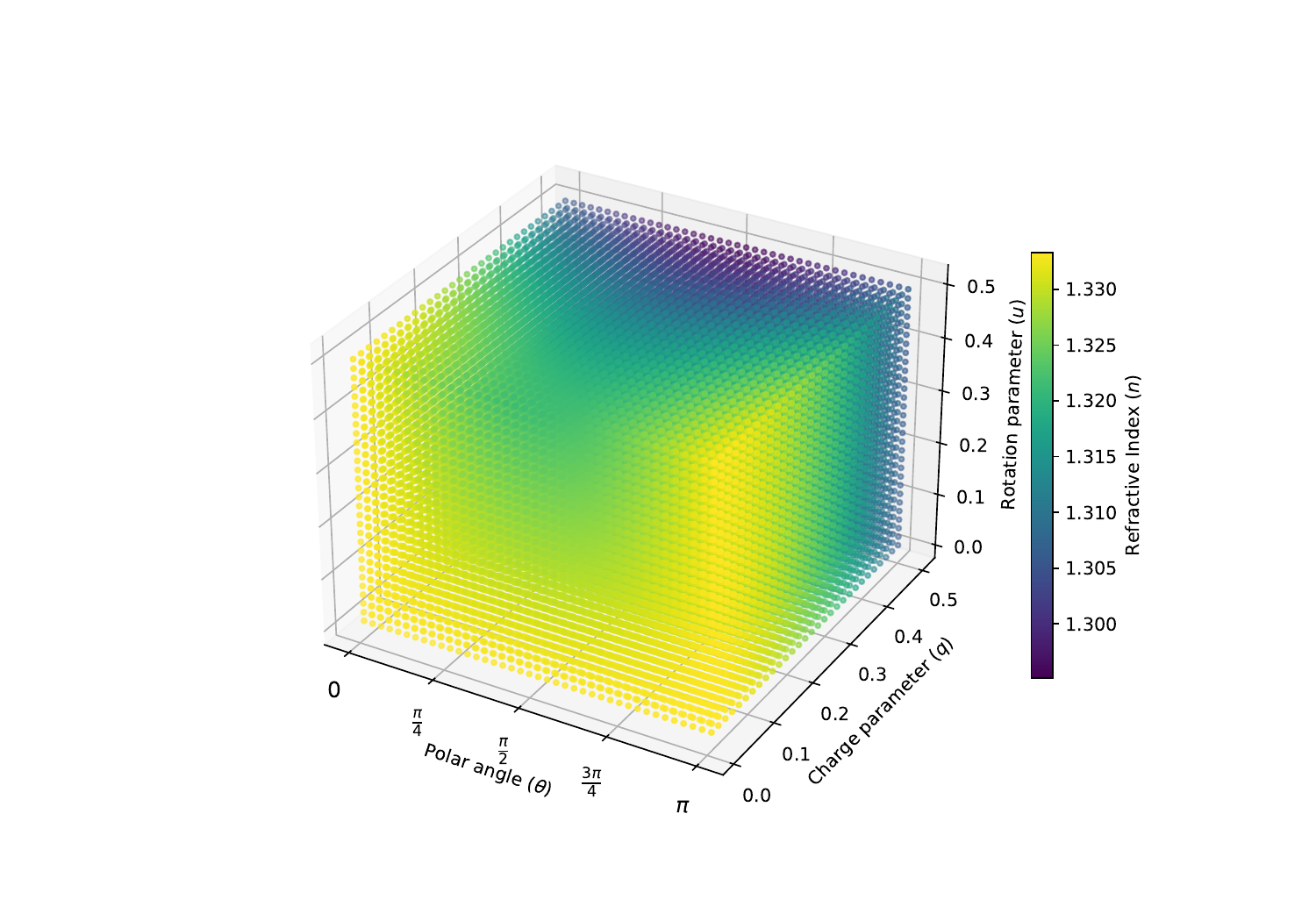}} 
    \end{subfigure}   
	\end{center}
	\caption{Three-dimensional scatter plots illustrating the dependence of the refractive index of the KNBH on the charge parameter, rotation parameter, and polar angle. Subfigure (a) corresponds to the prograde motion, while subfigure (b) represents the retrograde motion. Here we consider radial distance $x=10$. } \label{3DDfig}
\end{figure}

Fig. \ref{3DDfig} visualizes the variation of the refractive index with respect to the three key parameters: charge, rotation, and polar angle simultaneously. It serves as a comprehensive graphical summary of the preceding plots, enabling a unified understanding of how these parameters  collectively influence the optical behavior of the KNBH. In the prograde case, the refractive index exhibits an increasing trend with the rotation parameter and a decreasing trend with charge. The effect is most pronounced at equatorial plane $(\theta=\pi/2)$, where the refractive index attains its maximum, indicating a stronger deflection. Conversely, in the retrograde case, the refractive index is generally reduced. It decreases with increasing rotation and increases with charge, showing inverse behavior compared to the prograde scenario. Additionally, the minimum refractive index typically occurs near the polar regions $(\theta=0, \pi)$.\\

These plots collectively highlight the significant and nontrivial interplay between charge, rotation, and angular position in shaping the refractive index profile.

\section{Deflection due to Kerr-Newman Black Hole}
The propagation of light rays passing through a region of space having a gravitational field can be equated with a situation where the light ray is passing through an optical medium characterized by a refractive index at a point whose value depends upon the gravitational field at that point.  Thus, the curved space-time is related to a special optical medium with graded refractive index, and the light ray will deviate due to the variation of the refractive index of the corresponding media.

\begin{figure} [H]
    \begin{tikzpicture}[scale=2]

  \draw[->] (0,0,0) -- (2,0,0) node[anchor=north]{$y$-axis};
  \draw[->] (0,0,0) -- (0,2,0) node[anchor=north]{$z$-axis};
  \draw[->] (0,0,-3) -- (0,0,2) node[anchor=north]{$x$-axis};
  \node at (0.9,1.9,0.1) {\footnotesize $\phi = \pm \frac{\pi}{2}$};
  \fill[blue!10, opacity=0.6] (0,0,0) -- (2,0,0) -- (2,2,0) -- (0,2,0) -- cycle;
  \node at (0.7,0.1,1) {\footnotesize $x$-$y$ plane};

  \fill[green!10, opacity=0.8] (0,0,0) -- (2,0,0) -- (0,-2,-3) -- (0,0,2) -- cycle;
  \node at (0.9,0.2,-0.5) {\footnotesize $y$-$z$ plane};

\draw[-, dashed, blue] (1.2,2,0.7) -- (-0.5,0.41,1.5);
  \draw[->, thick, blue] (1.2,2,0.5) .. controls (-0.2,0.7,0.6) and (0.1,1,0.6) .. (-0.3,0.1,2);
   \node at (1.6,1.85,0.5) {\tiny Incoming light ray};
  \draw[-] (-0.7,-0.08,0.1) arc[start angle=245,end angle=155,radius=0.13]
   node[pos=1.7, anchor=east] {\footnotesize $\Delta \psi$};
      \node at (-0.9,-0.2,0.5) {\tiny Deflected light ray};
      \node at (-0.8,0.7,0.5) {\tiny Undeflected light ray};
  \draw[<->, thick, red] (1.2,2,1) -- (1.1,1.1,-0.2);
  \node at (1.5,2,1.7) {\footnotesize $b$};

  \shade[ball color=black] (0,0,0) circle (0.2);
  \draw[->, thick] (0.2,0.2,0.2) arc[start angle=0, end angle=125, radius=0.15];
  \node[text=white] at (0,0,0) {\footnotesize BH};

\draw[thick] (-1.2,-1.2,0) ellipse (0.15 and 0.08);
\fill (-1.2,-1.2) circle (0.03); 
\node at (-1.2,-1.5,0) {\footnotesize Observer};
\end{tikzpicture}
    \caption{Schematic diagram illustrating the deflection of light. }
    \label{defsch}
\end{figure}
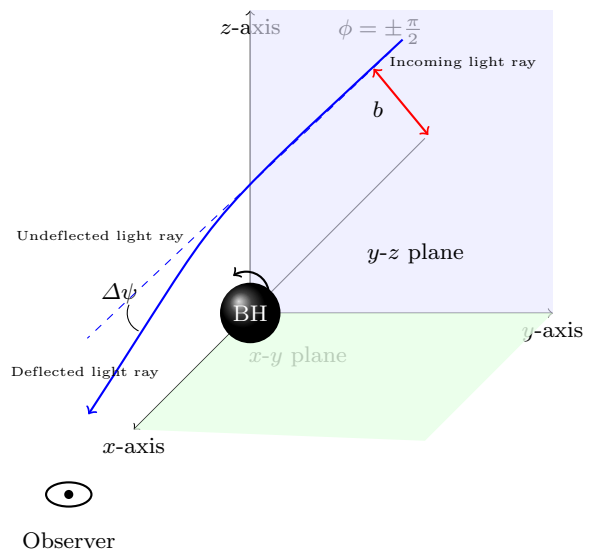

Now, the deflection of the light ray that depends on the refractive index of the medium can be represented as \cite{born2013principles, sen2010more, roy2015trajectory, roy2017deflection, roy2025deflection}, 
\begin{equation} \label{Eq_def}
    \Delta \psi=2\int_{b}^{\infty}\frac{dr}{r\sqrt{(\frac{n(r).r}{n(b).b})^2-1}}.
\end{equation}
In the present case, light approaches from asymptotic infinity, parallel to x-axis ($r =- \infty$ or $x =- \infty$) toward a charged rotating gravitational mass at the origin, defined by the Schwarzschild radius $r_g$, rotation parameter $\alpha$, and associated charge $Q$.
The beam reaches $r = \infty$ or $x = \infty$ after a specified amount of deflection ($\Delta\psi$). The closest approach distance to the incoming ray (known as the impact parameter) is the perpendicular distance between the center of the KNBH and the ray.  Now, in our consideration, all the rays are cutting the y-z plane perpendicularly. On the y-z plane, for all the points, x-coordinate is zero and when the ray cuts the plane, $\phi = -\pi/2$  or $+\pi/2$. So, the perpendicular distance between the ray and the center of the KNBH is $\sqrt{y^2 +z^2}=r=b$ (b is the physical radius of the KNBH), shown in Fig.~\ref{defsch}. 
\newpage
\begin{strip}
Now, we change the variable to $x = \frac{r}{r_g}$ so that $dr = r_gdx$ and the corresponding limit changes from $x = v$ to $x =\infty$, as the limit of $r$ changes from $r = b$ to $r =\infty$ in Eq.(\ref{Eq_def})
\begin{equation} \label{Eq_def_2}
\begin{split}
     \Delta \psi&=2\int_{v}^{\infty}\frac{dx}{x\sqrt{(\frac{n(x).x}{n(v).v})^2-1}}-\pi =2I-\pi,
\end{split}
\end{equation}
where 
\begin{equation}\label{Eq_I}
\begin{split}    
    I&=\int_{v}^{\infty}\frac{dx}{x\sqrt{(\frac{n(x).x}{n(v).v})^2-1}}\\
    &= n(v).v\int_{v}^{\infty}\frac{dx}{x\sqrt{(n(x).x)^2-(n(v).v)^2}}.
\end{split}
\end{equation}
Substituting the value of the refractive index from Eq.(\ref{Eq_ref_x_FD}), we rewrite the above Eq. (\ref{Eq_I}) of $I$  as

\begin{equation} 
\begin{split}
    I &=  D_{KN}\int_{v}^{\infty}\frac{dx}{x\sqrt{[n_0(x).x(1+C_x)^{-1}(1+2S_x)^{-\frac{1}{2}}]^2-D^2_{KN}}} \\
    &=D_{KN}\int_{v}^{\infty}\frac{dx}{x\sqrt{n^2_0(x).x^2(1+C_x)^{-2}(1+2S_x)^{-1}-D^2_{KN}}} \\
    &=D_{KN}\int_{v}^{\infty}\frac{dx}{x\sqrt{n^2_0(x).x^2-D^2_0+n^2_0(x).x^2(1+C_x)^{-2}(1+2S_x)^{-1}-n^2_0(x).x^2+D^2_0-D^2_{KN}}} \\
    &=D_{KN}\int_{v}^{\infty}\frac{dx}{x\sqrt{n^2_0(x).x^2-D^2_0}}\left[1+\frac{n^2_0(x).x^2
    \{(1+C_x)^{-2}(1+2S_x)^{-1}-1\}+D^2_0-D^2_{KN}}{n^2_0(x).x^2-D^2_0}\right]^{-\frac{1}{2}} \\
    &=D_{KN}\int_{v}^{\infty}\frac{dx}{x\sqrt{n^2_0(x).x^2-D^2_0}}[1+P(x)]^{-\frac{1}{2}},
\end{split}
\end{equation} 
with $D_{0}=n_{0}(v).v$, corresponding to SBH deflection, $n(v).v = D_{KN} = D_0(1+C_v)^{-2}(1+2S_v)^{-1}$ and also we have denoted
\begin{equation}
    P(x)= \frac{n^2_0(x).x^2\{(1+C_x)^{-2}(1+2S_x)^{-1}-1\}+D^2_0-D^2_{KN}}{n^2_0(x).x^2-D^2_0}.
\end{equation}
Eq. (\ref{Eq_def_2}) may be used to obtain the exact deflection of light on the non-equatorial plane of a KNBH as
\begin{equation} \label{Eq_def_final}
\begin{split}
     \Delta \psi&=2I-\pi\\
     &=2D_{KN}\int_{v}^{\infty}\frac{dx}{x\sqrt{n^2_0(x).x^2-D^2_0}}[1+P(x)]^{-\frac{1}{2}} -\pi.
\end{split}
\end{equation}
The above deflection angle can be expanded as a series
\begin{equation}
\label{Eq_def_final1}
\begin{split}
    \Delta \psi =&2D_{KN}\int_{v}^{\infty}\frac{dx}{x\sqrt{n^2_0(x).x^2-D^2_0}}\left[1-\frac{1}{2}P(x)  +\frac{3}{8}P^2(x)-\frac{5}{16}P^3(x)+ .................\right] -\pi.
\end{split}
\end{equation}
Also, the deflection of light ray in the equatorial plane can be obtained by considering the refractive index expressed by Eq. (\ref{Eq_ref_x_FD_equa}).
\end{strip}
\begin{figure} 
	\begin{center}
    \begin{subfigure}[]
     {\includegraphics[width=0.35\textwidth,height=0.19\textheight]{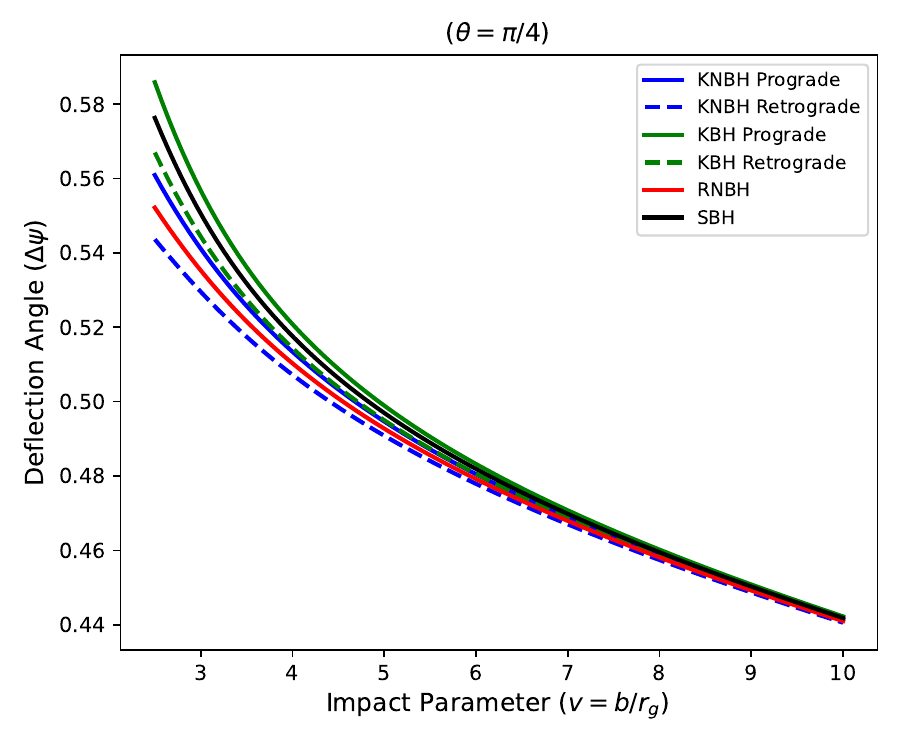}}   
    \end{subfigure}
     \begin{subfigure}[]
     {\includegraphics[width=0.35\textwidth,height=0.19\textheight]{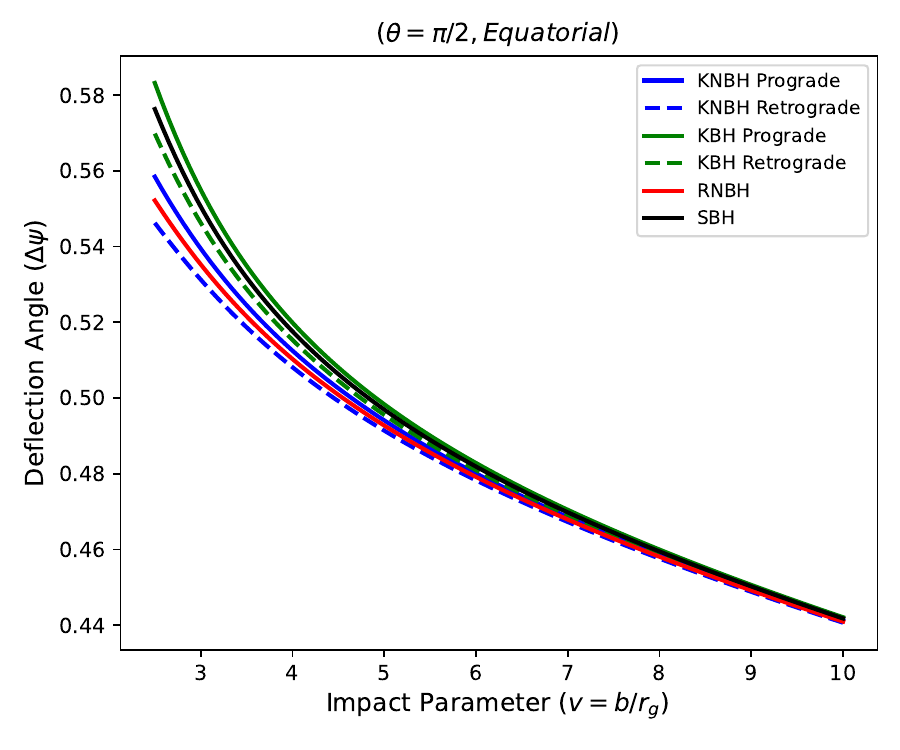}} 
    \end{subfigure}  
    \begin{subfigure}[]
     {\includegraphics[width=0.35\textwidth,height=0.19\textheight]{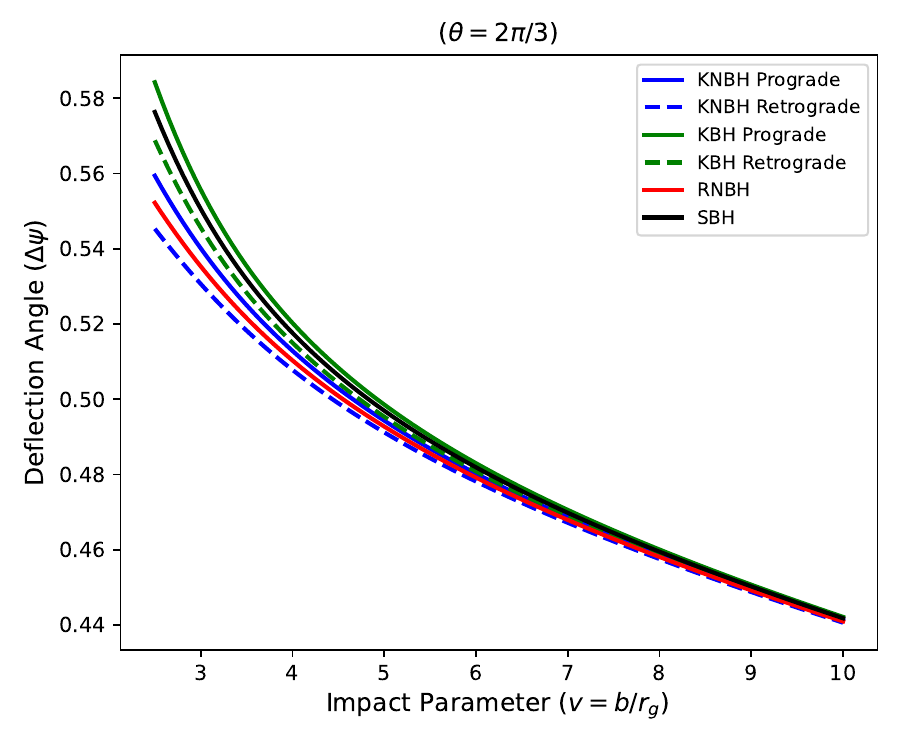}}  
    \end{subfigure}
     \begin{subfigure}[]
     {\includegraphics[width=0.35\textwidth,height=0.19\textheight]{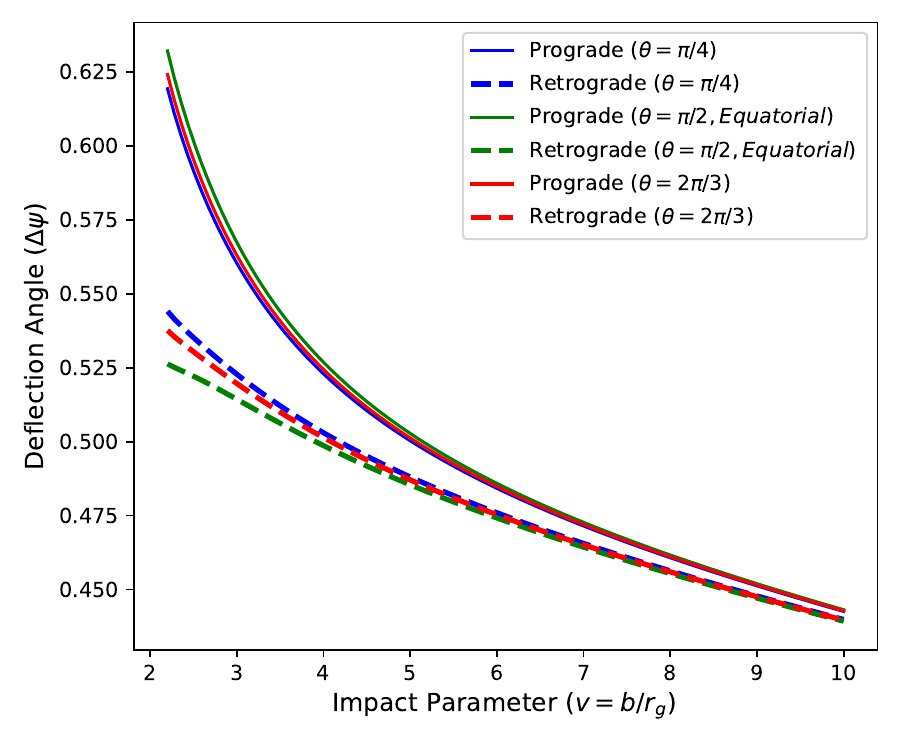}} 
    \end{subfigure}  
	\end{center}
	\caption{The variation of deflection angle with impact parameter. The plots also show the results for the SBH, RNBH and KBH for comparison. Here we consider $M=1$, $u=0.4$, $q=0.5$ and $x=10$.} \label{daip}
\end{figure}
\begin{figure} 
	\begin{center}
    \begin{subfigure}[]
     {\includegraphics[width=0.35\textwidth,height=0.19\textheight]{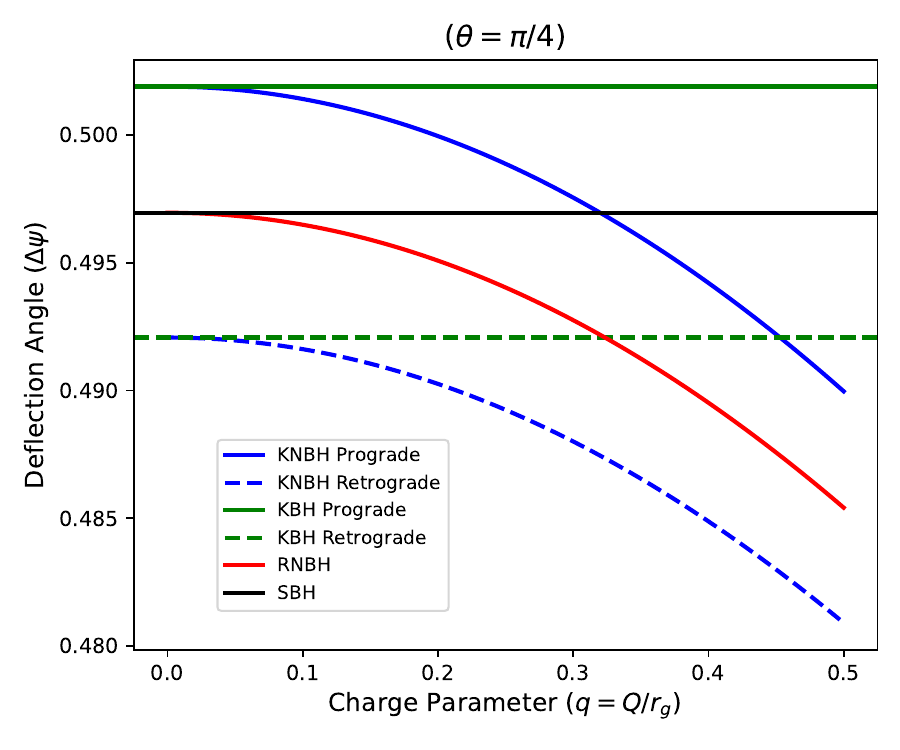}}   
    \end{subfigure}
     \begin{subfigure}[]
     {\includegraphics[width=0.35\textwidth,height=0.19\textheight]{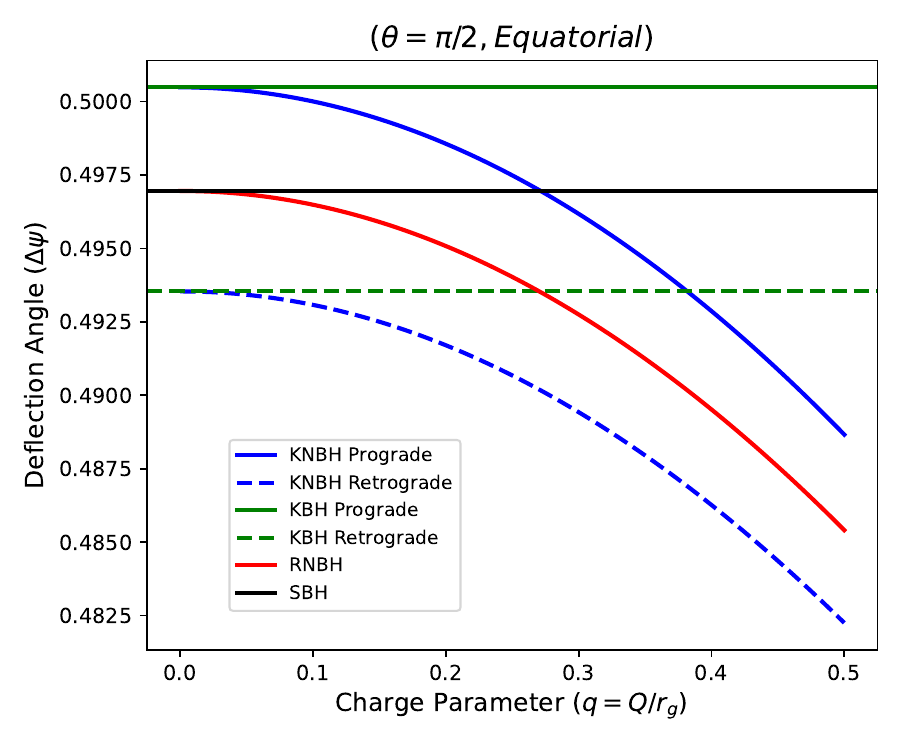}} 
    \end{subfigure}  
    \begin{subfigure}[]
     {\includegraphics[width=0.35\textwidth,height=0.19\textheight]{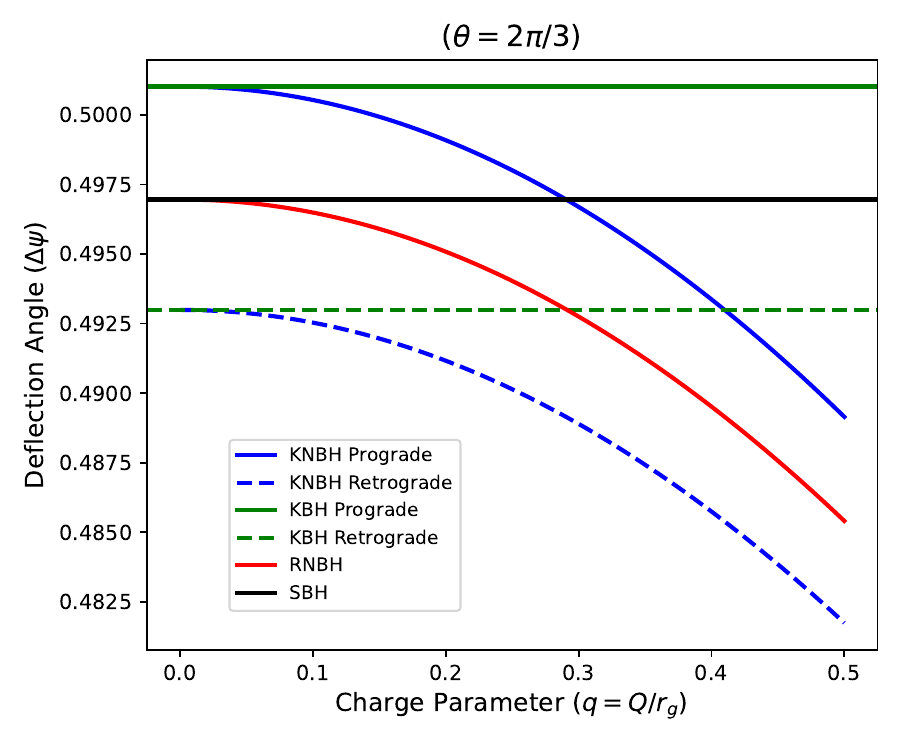}}   
    \end{subfigure}
     \begin{subfigure}[]
     {\includegraphics[width=0.35\textwidth,height=0.19\textheight]{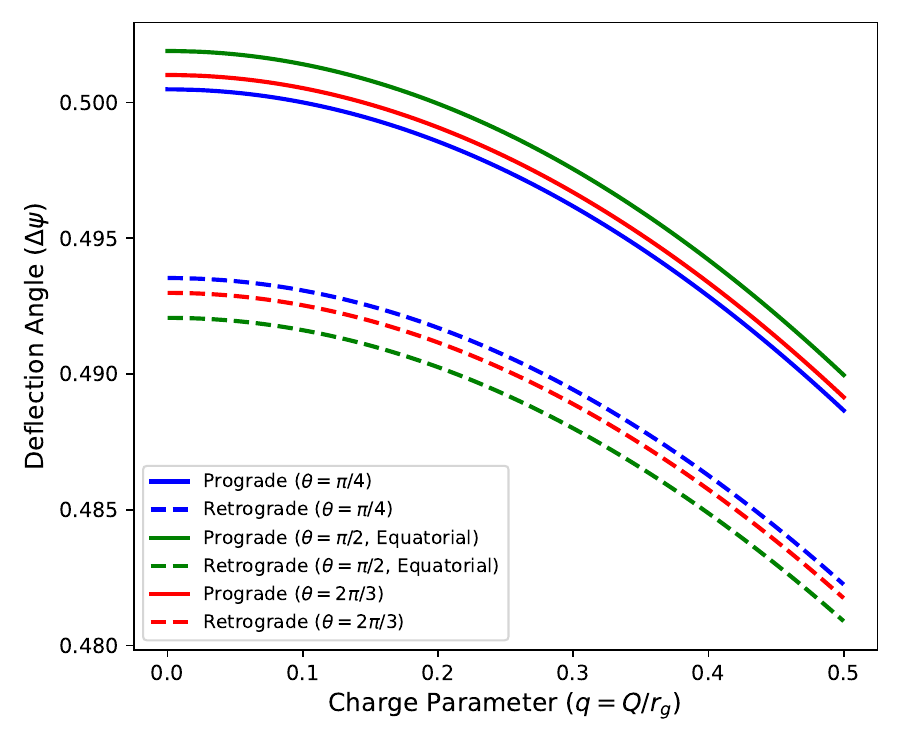}} 
    \end{subfigure}  
	\end{center}
	\caption{The variation of deflection angle with retrograde and prograde motion with varying charge parameter. Here we consider $M=1$, $u=0.4$, $q=0.5$ and $x=10$.} \label{DAQ}
\end{figure} 
\begin{figure} 
	\begin{center}
    \begin{subfigure}[]
     {\includegraphics[width=0.35\textwidth]{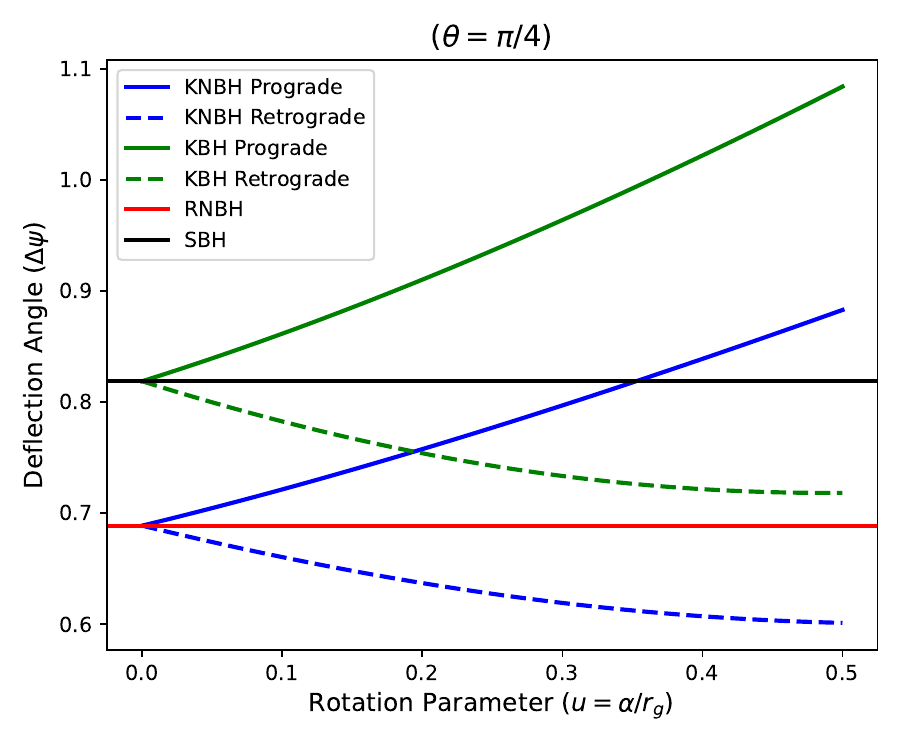}}   
    \end{subfigure}
     \begin{subfigure}[]
     {\includegraphics[width=0.35\textwidth]{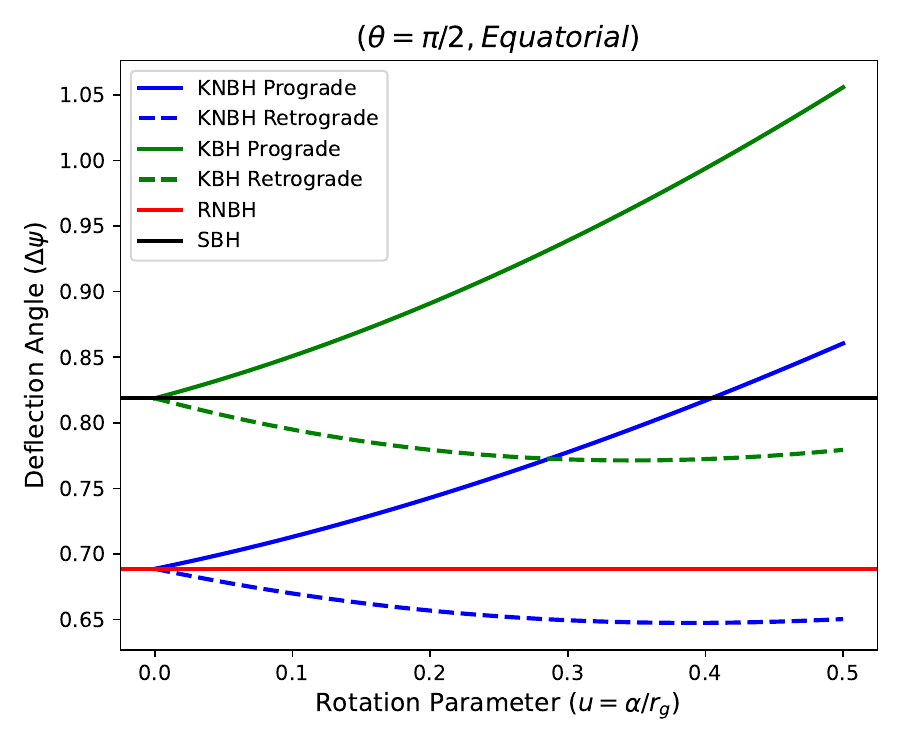}} 
    \end{subfigure}  
    \begin{subfigure}[]
     {\includegraphics[width=0.35\textwidth]{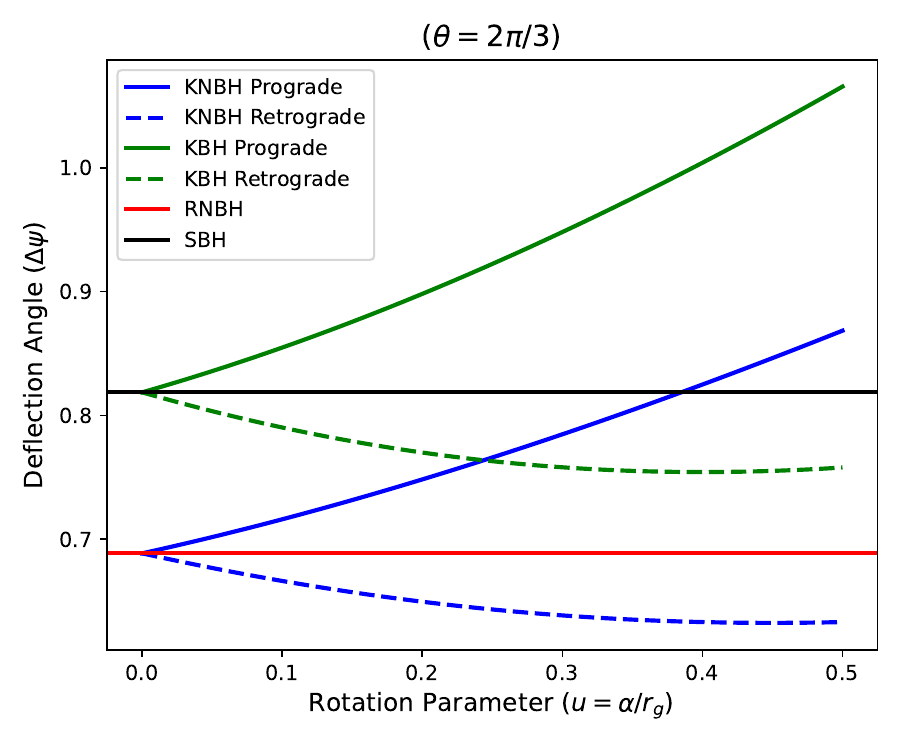}}   
    \end{subfigure}
     \begin{subfigure}[]
     {\includegraphics[width=0.35\textwidth]{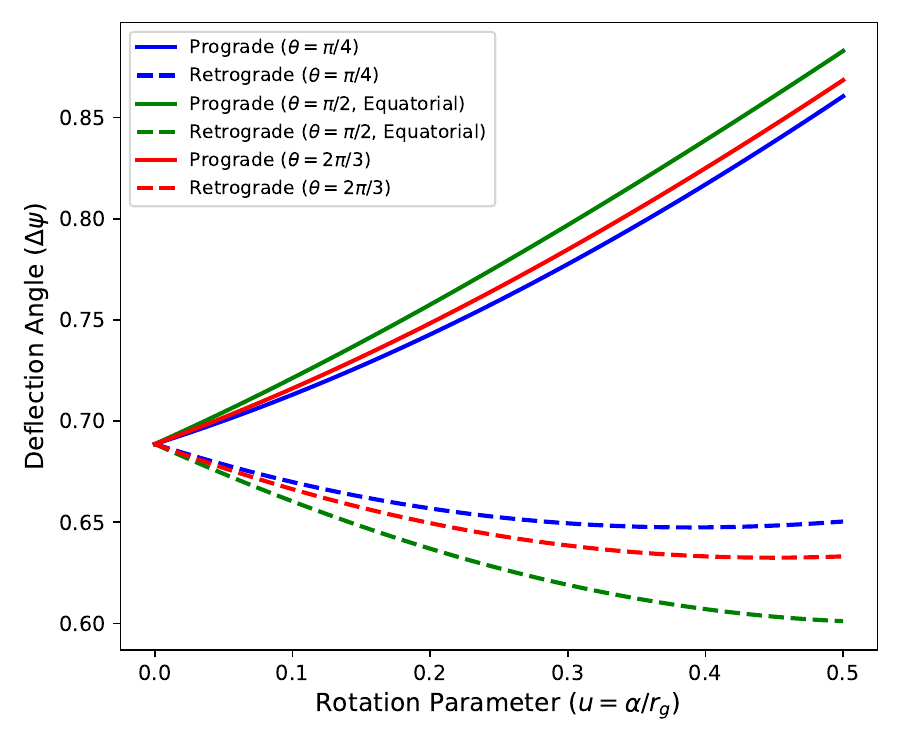}} 
    \end{subfigure}  
       \end{center}
	\caption{The variation of deflection angle with retrograde and prograde motion with varying rotation parameter. Here we consider $M=1$, $u=0.4$, $q=0.5$ and $x=10$.} \label{DAa}
\end{figure} 
Figs \ref{daip}-\ref{DAa}, collectively illustrate the variation of the deflection angle as a function of the impact parameter, charge parameter, rotation parameter and polar angle for KNBH and other BH solutions in GR. Fig. \ref{daip} shows that the deflection angle decreases monotonically with increasing impact parameter. 
Among the configurations, the deflection is most pronounced for the KBH in the prograde case and is least for the KNBH retrograde case.
The physical significance of this result is that the deflection angle is largest for the KBH in the prograde case due to frame dragging aiding the photon’s approach, increasing spacetime curvature. Conversely, it is smallest for the KNBH retrograde case, where opposing frame dragging and charge both reduce the effective bending. Furthermore,  the deflection angle for KNBH attains its maximum value on the equatorial plane for prograde motion and its minimum in the retrograde configuration. A comparative analysis with other BHs reveals the ordering of the deflection strength as: KBH prograde $>$ SBH $>$ KBH retrograde $>$ KNBH prograde $>$ RNBH  $>$ KNBH retrograde. Fig. \ref{DAQ} depicts the variation of the deflection angle with the charge parameter. Since the KBH and SBH are independent of charge, their deflection angles remain constant and are represented by horizontal lines. From the graphical representation, it is once again evident that the KBH prograde configuration exhibits the most dominant deflection, while the KNBH retrograde case shows the least. For the KNBH alone, the deflection is more pronounced in the prograde configuration compared to the retrograde one, and it also exceeds that of the RNBH. Further, the maximum deflection is observed for the prograde motion at the equatorial plane and the minimum for the retrograde motion in the same plane. Fig. \ref{DAa} presents the dependence of the deflection angle on the rotation parameter. In this case, the RNBH and SBH yield constant deflection angles due to the absence of rotational dependence, appearing as horizontal lines. For the KNBH, the deflection angle increases with rotation in the prograde case and decreases in the retrograde case. In addition, inducing charge parameters lead to a reduction in the deflection angle for both cases. Comparatively, the KBH (uncharged rotating BH) exhibits a stronger deflection than the KNBH, underscoring the diminishing influence of charge on the deflection angle. The dependency of polar angle on deflection angle remains consistent with previous cases, having reached a maximum on the equatorial plane and decreasing toward the polar plane.
\section{Comparison with Kerr-Sen Black Hole}
\begin{figure} 
	\begin{center}
    \begin{subfigure}[]
     {\includegraphics[width=0.35\textwidth]{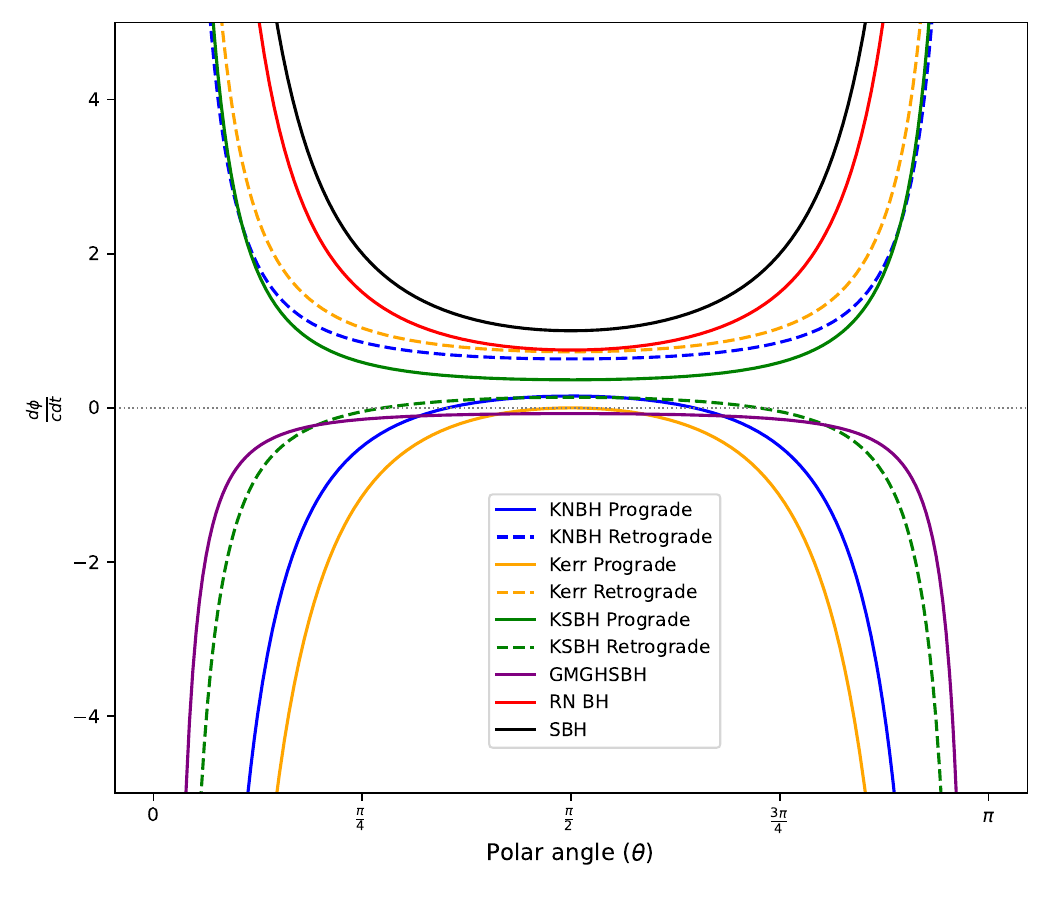}} 
    \end{subfigure}
     \begin{subfigure}[]
     {\includegraphics[width=0.4\textwidth]{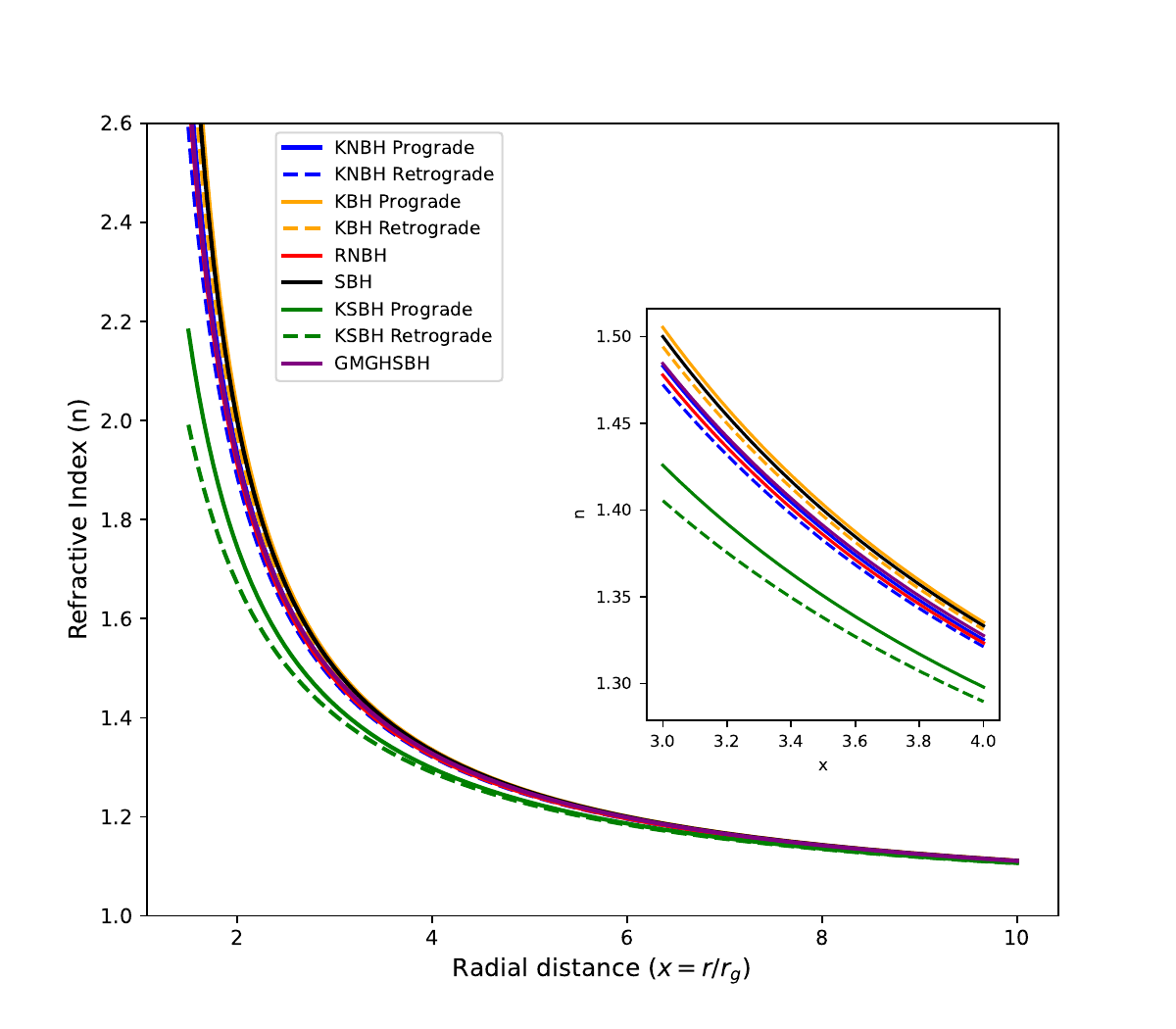}} 
    \end{subfigure}
     \begin{subfigure}[]
     {\includegraphics[width=0.4\textwidth]{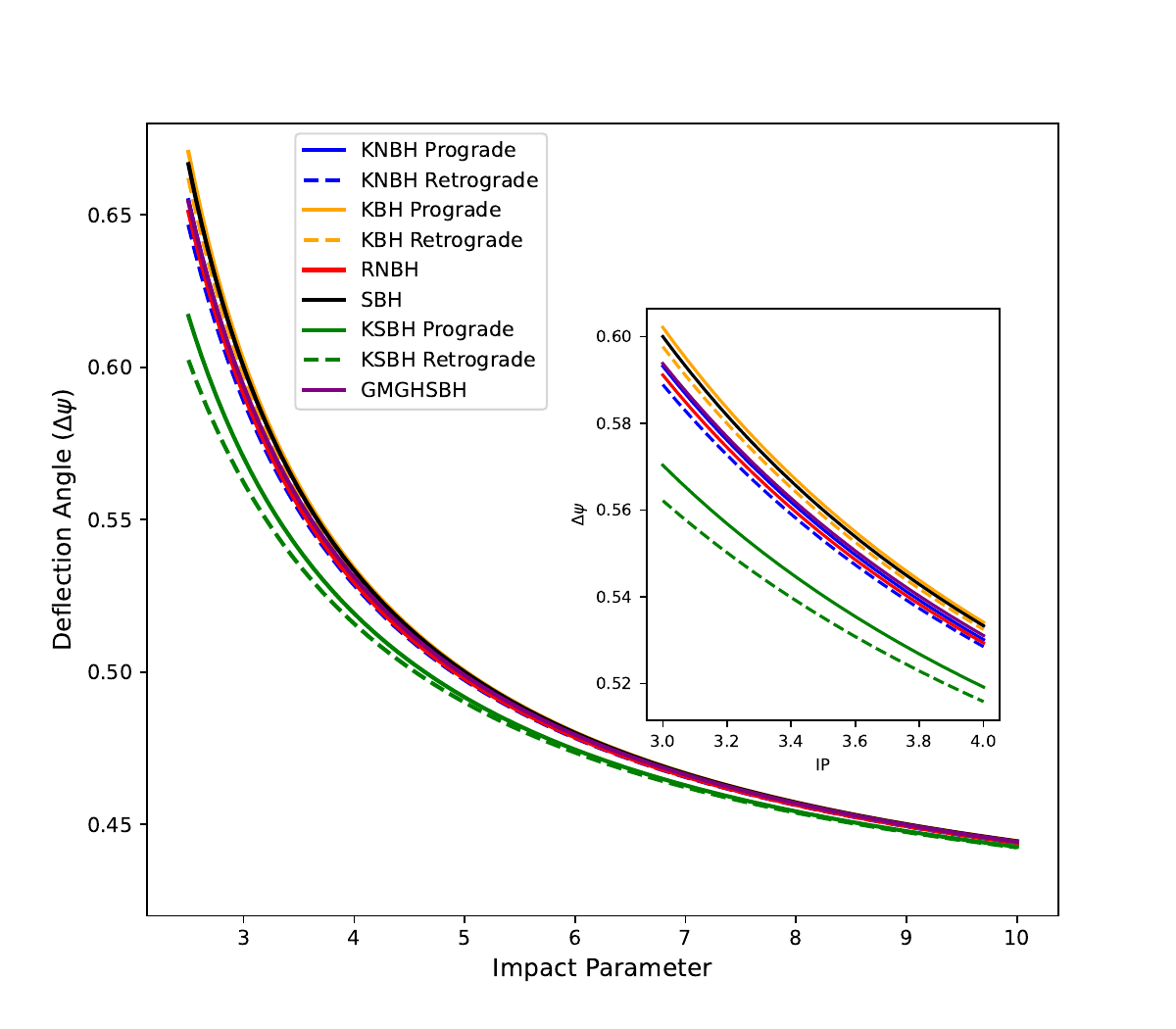}} 
    \end{subfigure}
	\end{center}
	\caption{Graphical comparison between the KSBH and KNBH spacetimes, along with the corresponding BH solutions to which they reduce in specific limits. (a) Variation of frame dragging with the polar angle, here we consider $M=1$, $E_0=-0.5$, $L=\pm0.5(Retro/Pro)$, $x=10$, $u=0.4$ and $q=0.5$. (b) Variation of the refractive index with radial distance, here we consider $M=1$,  $v=\pm4(Retro/Pro)$, $\theta=\pi/2$, $u=0.4$ and $q=0.5$. (c) Variation of deflection angle with impact parameter, here we consider $M=1$, $\theta=\pi/2$, $u=0.4$ and $q=0.5$.} \label{KNKSCompfig}
\end{figure}
The spacetime characterized by the Kerr-Sen metric is not vacuum, similar to the Kerr-Newman situation in Einstein-Maxwell theory. Although both matrices describe rotating and charged BH space times, however, the Kerr-Sen case is due to modifications of GR coming from the low energy heterotic string theory effective action. 
In this section, we present a comparison between KSBH and KNBH according to our recent result with respect to KSBH \cite{roy2025deflection}. Fig. \ref{KNKSCompfig} represents the graphical comparison between the KSBH and KNBH space times. From Fig. \ref{KNKSCompfig}(a) it is observed that KSBH shows the opposite behaviour with respect to other well-known BH solutions when trajectories are compared in terms of frame dragging. Also, from Figs. \ref{KNKSCompfig}(b) and (c) it is clear that the refractive index and hence the deflection angle of light due to KSBH is the least whereas for KBH is highly pronounced by keeping the KNBH as intermediate. Although both KSBH and KNBH changes to KBH by vanishing the dilation field and charge, respectively, the string frame seems to be more fundamental than the Einstein one, as some hidden symmetries are present in the Kerr-Sen spacetime. 
It is observed that the KNBH prograde deflection is more pronounced compared to the KSBH prograde; a similar trend is noted for the retrograde case. Thus, KNBH prograde shows the highest deflection, while KSBH retrograde exhibits the lowest. The greater deflection in KNBH arises because the dilaton–axion field in KSBH weakens the effective gravitational field \cite{sen1992rotating, Horowitz:1992jp}.
\section{Summary, Conclusions and Future directions}
In this work, we study the non-equatorial deflection angle of light (approaching parallel to the equatorial plane and x-axis) around a KNBH, using the material medium approach and compare its behavior with other well-known solutions in GR, such as KBH, RNBH and SBH. To the best of our knowledge, we are the first to incorporate the non-equatorial deflection angle for KNBH. The obtained results are classified based on the photon sphere, frame dragging, and deflection angle, and are summarized as follows:\\\\
\textbf{Photon Sphere}
\begin{itemize} [label=$\bullet$]
    \item In the polar plane, the radius of the photon sphere decreases with increasing charge and rotation. It is evident that at the polar plane the SBH possesses the largest photon sphere radius, whereas the KNBH with higher charge and rotation exhibits the smallest photon sphere radius.
    \item However, in the equatorial plane, for the prograde case, the photon sphere radius decreases with increasing charge and rotation, whereas for the retrograde case, it increases with charge and rotation. At higher rotations and low charge values, the photon sphere vanishes.
\end{itemize}
\textbf{Frame-Dragging}
\begin{itemize} [label=$\bullet$]
    \item The frame dragging effect is more prominent in prograde motion than in retrograde motion with charge and rotation, which means that in the presence of spin on the BH in the same direction as the light ray, the drag is greater than in the opposite direction of the light ray.
    \item In the absence of charge and spin,  the convergent nature of the trajectories reveals the prograde and retrograde nature of the BH. Depending on the values of the rotation parameter and charge parameter, the frame dragging effect is the same at some point, for both prograde and retrograde motion under KNBH and this nontrivial nature results because of the interplay between charge and rotation.
    \item Another important fact to note is that frame dragging increases towards poles for retrograde trajectories while decreasing for prograde trajectories in KNBH geometry. The same effect is also observed in the case of all geometries in GR including KSBH of string theory. 
    \item The magnitude of frame dragging is higher near the pole surface than the surface of the equatorial plane. This is due to the rotation frequency of the BH for which the pole is closer to the center than the equatorial surface. This result shows excellent agreement with Chakraborty et al. \cite{chakraborty2014dragging}.
    \item It is also observed that KSBH shows the opposite behaviour with respect to other well-known BH solutions when trajectories are compared in terms of the frame dragging.
\end{itemize}
\textbf{Deflection Angle}
\begin{itemize} [label=$\bullet$]
    \item The deflection angle decreases with increasing impact parameter and charge but increases with rotation in the prograde case and decreases in the retrograde case.
    \item The comparative analysis suggests that the KNBH prograde configuration exhibits intermediate bending behavior—greater than that of the RNBH and, but generally lower than that of the KBH retrograde, SBH and KBH prograde respectively. In contrast, the KNBH retrograde case results in the lowest deflection. Physically, this can be summarized as follows: the combination of favorable frame dragging (prograde) and the absence of charge (as in KBH) leads to maximum deflection, while opposing frame dragging (retrograde) combined with the charge-induced weakening of gravity (as in KNBH) results in minimal deflection.
    \item Notably, KNBH shows strong angular dependence, with maximum deflection in the equatorial plane for prograde motion and minimum towards the poles for retrograde motion, distinguishing it from charge and spin independent cases.
    \item For the equatorial plane of KNBH, the obtained results show excellent agreement with those of You-Wei Hsiao et al. \cite{hsiao2020equatorial}. Furthermore, setting the charge parameter to zero, the KNBH reduces to the KBH; by setting the rotation parameter to zero, the KNBH reduces to the RNBH; and the results align perfectly with the findings of Roy et al. \cite{roy2015trajectory,roy2017deflection}. Also, setting both the charge and rotation parameters to zero, the KNBH reduces to the SBH and this result perfectly agrees with Sen \cite{sen2010more}.
   \item We also compared the deflection angle results obtained for the equatorial plane case with the KSBH recently presented by Roy et al. \cite{roy2025deflection}. It is observed that the KNBH prograde deflection is more pronounced compared to the KSBH prograde; a similar trend is noted for the retrograde case. Thus, KNBH prograde shows the highest deflection, while KSBH retrograde exhibits the lowest. The greater deflection in KNBH arises because the dilaton–axion field in KSBH weakens the effective gravitational field.
   \item Finally, we compare our findings with those of Wang et al. \cite{Wang:2022ouq}, who obtained KNBH geodesics in terms of elliptic integrals and Jacobian elliptic functions for both null and timelike trajectories, using the Mino time formalism with specified initial conditions. Their analysis reveals boomerang-like spherical orbits arising from frame dragging and black hole charge. Our results exhibit similar qualitative trends, where the deflection decreases with increasing charge and impact parameter, increases with spin in prograde motion, and decreases in retrograde motion, consistent with their frame-dragging interpretation. The angular dependence we obtain, strongest in the equatorial plane and gradually diminishing toward the poles, also aligns with their conclusions. However, as the two studies employ different formalisms, a direct quantitative comparison is not possible and the agreement should be regarded as qualitative.
\end{itemize}
\textbf{Future Directions}
\begin{itemize} [label=$\bullet$]
\item It is well established that extremal BHs exhibit distinctive features in their photon sphere structure and light deflection behavior. In particular, as demonstrated by Chen et al. \cite{Chen:2024oyv}, the extremal KNBH permits a stable photon orbit exactly at the horizon due to the presence of a degenerate root in the radial potential. This leads to a qualitatively different behavior in the deflection angle compared to subextremal configurations, including strong deviations near the critical impact parameter. These properties are unique to the extremal limit and influence photon trajectories in ways not observed in non-extremal BHs. Although the present study does not consider the extremal regime, it would be interesting in future work to investigate such cases using the material medium approach.
\end{itemize}
\begin{acknowledgements}
We sincerely thank the anonymous referee for their careful review and constructive comments, which have improved the scientific content and overall quality of this paper.
Authors, SR and PG express sincere and deep gratitude to the Department of Physics, NITA, for providing the necessary research environment to complete this work. 
The author, SK, sincerely acknowledges IMSc for providing exceptional research facilities and a conducive environment that facilitated his work as an Institute Postdoctoral Fellow. One of the authors, HN, would like to thank IUCAA, Pune, for the support under its associateship program, where a part of this work was done. The author HN also acknowledges the financial support provided by the Science and Engineering Research Board (SERB), New Delhi, through grant number CRG/2023/008980.
\end{acknowledgements}

\appendix
\renewcommand\theequation{\Alph{section}\arabic{equation}} 
\setcounter{section}{1} 
\setcounter{equation}{0}
\section*{Appendix A}

The expression for the photon sphere on the equatorial plane of the KNBH is given in Eq.  \eqref{rpheq2}. The detailed form of the symbols involved in the equation is provided below.
\begin{equation}
    \mathcal{A} = \frac{1}{2} \sqrt{ -3M^2-\frac{8}{3}Q^2 + \frac{1}{3} \left( \mathcal{B} + \frac{\Delta_0}{\mathcal{B}} \right) },
\end{equation}
where
\begin{equation}
    \mathcal{B} = \sqrt[3]{\frac{1}{2} \left( \Delta_1 + \sqrt{ \Delta_1^2 - 4\Delta_0^3 } \right)},
\end{equation}
with
\begin{equation}
    \Delta_0 = 81M^4 - 72M^2\alpha^2 - 144M^2Q^2 + 48Q^2\alpha^2 + 64Q^4,
\end{equation}
and 
\begin{align}
\Delta_1 &= 2(9M^2 + 4Q^2)^3 
- 216M(9M^2 + 4Q^2)(\alpha^2 + 3Q^2) \notag\\
&\quad + 3888M^2Q^2(\alpha^2 + Q^2) 
+ 432M^2(\alpha^2 + 3Q^2)^2 \notag\\
&\quad - 288Q^2(\alpha^2 + Q^2)(9M^2 + 4Q^2).
\end{align}
    

\bibliography{KerrNewmann}
\bibliographystyle{unsrt}

\end{document}